\documentclass[useAMS,usenatbib]{mn2e}
\usepackage{amssymb,amsmath,graphicx}
\usepackage{longtable}[1]
\usepackage{graphics}
\usepackage{threeparttable}
\usepackage{lscape}
\usepackage{color}
\usepackage{enumitem}
\usepackage{xspace} 
\usepackage{url}
\usepackage{float}
\usepackage[none]{hyphenat}

\newcommand{\Ha}{$\mathrm{H}\alpha$\xspace}
\newcommand{\NII}{$[\mathrm{N}\textsc{ii}]$\xspace}
\newcommand{\NIIa}{$[\mathrm{N}\textsc{ii}]\,\lambda 6548$\xspace}
\newcommand{\NIIb}{$[\mathrm{N}\textsc{ii}]\,\lambda 6583$\xspace}
\newcommand{\FHa}{${F}_{\mathrm{H}\alpha}$\xspace}
\newcommand{\FNII}{${F}_{[\mathrm{N}\textsc{ii}]}$\xspace}
\newcommand{\WHa}{${W}_{\mathrm{H}\alpha}$\xspace}

\newcommand{\ratio}{$[\mathrm{N}\textsc{ii}]\,/\,\mathrm{H}\alpha$\xspace}

\title[OMEGA II. -- Environmental influence on integrated star formation properties and AGN activity]{OMEGA -- OSIRIS Mapping of Emission-line Galaxies in A901/2: II. -- Environmental influence on integrated star formation properties and AGN activity}
\author[Rodr\'iguez del Pino et. al.]{Bruno Rodr\'iguez del Pino$^{1,2}$\thanks{E-mail: brodriguez@cab.inta-csic.es}, 
Alfonso Arag\'on-Salamanca$^{2}$,
Ana L. Chies-Santos$^{2,3,4}$, 
\newauthor
Tim Weinzirl$^{2}$,
Steven P. Bamford$^{2}$,
Meghan E. Gray$^{2}$,
Asmus B\"ohm$^{5}$,
\newauthor
Christian Wolf$^{6}$, David T. Maltby$^{2}$\\
$^{1}$Centro de Astrobiolog\'ia, INTA-CSIC, Torrej\'on de Ardoz, 28850, Madrid, Spain \\
$^{2}$School of Physics and Astronomy, The University of Nottingham, University Park, Nottingham, NG7 2RD, UK\\
$^{3}$Departamento de Astronomia, Instituto de F\'isica, Universidade Federal do Rio Grande do Sul, Porto Alegre, R.S, Brazil\\
$^{4}$Instituto de Astronomia, Geof\'isica e Ci\^encias Atmosf\'ericas, Universidade de S\~ao Paulo, S\~ao Paulo, SP, Brazil\\
$^{5}$Institute for Astro- and Particle Physics, University of Innsbruck, Technikerstr. 25/8, 6020 Innsbruck, Austria\\
$^{6}$Research School of Astronomy and Astrophysics, Australian National University, Cotter Road, Weston Creek, ACT 2611, Australia\\
}

\begin{document}

\date{Accepted for publication in MNRAS.}

\pagerange{\pageref{firstpage}--\pageref{lastpage}} \pubyear{2016}

\maketitle

\label{firstpage}

\begin{abstract}

We present a study of the star formation and AGN activity for galaxies in the Abell 901/2 multi-cluster system at $z\sim0.167$ as part of the OMEGA survey. Using Tuneable Filter data obtained with the OSIRIS instrument at the GTC we produce spectra covering the \Ha and \NII spectral lines for more than 400 galaxies. Using optical emission-line diagnostics, we identify a significant number of galaxies hosting AGN, which tend to have high masses and a broad range of morphologies. Moreover, within the environmental densities probed by our study, we find no environmental dependence on the fraction of galaxies hosting AGN. The analysis of the integrated \Ha emission shows that the specific star formation rates (SSFRs) of a majority of the cluster galaxies are below the field values for a given stellar mass. We interpret this result as evidence for a slow decrease in the star formation activity of star-forming galaxies as they fall into higher-density regions, contrary to some previous studies which suggested a rapid truncation of star formation. We find that most of the intermediate- and high-mass spiral galaxies go through a phase in which their star formation is suppressed but still retain significant star-formation activity. During this phase, these galaxies tend to retain their spiral morphology while their colours become redder. The presence of this type of galaxies in high density regions indicates that the physical mechanism responsible for suppressing star-formation affects mainly the gas component of the galaxies,  suggesting that ram-pressure stripping or starvation are potentially responsible.

\end{abstract}


\section{Introduction}

Despite significant advances in our understanding of galaxy evolution there is still ongoing debate on to what extent the fate of galaxies is set by their intrinsic properties, such as their mass (`nature'), or by the environment where they reside (`nurture'). It is clear now that both mass and environment act as strong drivers of galaxy evolution through cosmic time and as such they are connected to many galaxy properties. 

The morphologies, colours and star formation rates (SFRs) of galaxies change as a function of both stellar mass and environmental density \citep{Dressler_1980, Gomez_2003,Goto_2003,Kauffmann_2003_sfhs,Blanton_2005,Bamford_2009,Vulcani_2011}. 
Of particular interest is the correlation between SFR and stellar mass, usually referred to as the `star formation main sequence' \citep{Salim_2007, Whitaker_2012, Lee_2015}. Stellar mass also imposes a limit above which internal processes lead to the cessation of star formation \citep{Baldry_2006}. In general, quenching of star-formation can be produced by the establishment of a virialized hot halo (e.g., \citealt{Rees_1977}), leading to longer cooling times; star formation can be further prevented by feedback from supernovae and active galactic nuclei (AGN; \citealt{Booth_2009}, \citealt{Newton_2013}). The formation and sustaining of AGN are subject to the availability of gas acting as fuel. The amount of available gas to fuel the AGN might also be affected by the environment \citep{Chung_2009, Yara_2015}, which can strip or deplete the gas content of the galaxies. In fact, the presence of AGN in galaxies has been observed to be somehow related to the environment: while low-luminosity AGN do not show particular preference for any kind of environment \citep{Martini_2002,Miller_2003,Martini_2006}, luminous AGN are preferentially found in the field \citep{Kauffmann_2004} and groups \citep{Popesso_2006} rather than in clusters. In a more recent work \citep{desouza_2016}, the latter trend has been found to hold for elliptical galaxies but not for spiral ones. This whole picture is further complicated by the existent link between mass and environment, as more massive galaxies tend to reside in high-density regions \citep{Bolzonella_2010}, although \citet{Peng_2010} showed that their effects can be separated up to $z\sim1$. 

As a result of these correlations there is a clear bimodality in the observed properties of galaxies. Galaxies living in low-density regions tend to have low masses, blue colours, late-type morphologies and are actively forming stars, populating what is called the \textit{blue cloud} in the colour-magnitude diagram \citep{Strateva_2001}. In contrast, high-density regions tend to be populated by more massive, red objects with early-type morphologies and low or absent star formation, occupying the so-called \textit{red sequence} in the colour-magnitude diagram. 

The existence of this bimodality indicates that objects falling into higher density regions must experience a suppression of their star formation, accompanied by a change in their colours and, eventually, their morphologies. The low number of objects inhabiting the \textit{green valley} (region between the \textit{blue cloud} and the \textit{red sequence} in the colour-magnitude diagram) implies that the transition from blue to red must be quite rapid. Such a fast transition is supported by the substantial change in the fractions of blue and red galaxies as a function of environment, whereas the colours and \Ha equivalent widths of  star-forming objects remain relatively invariant \citep{balogh_2004_colour, Bamford_2008}. However, the timescales associated with the change in SFR and the morphological transformation seem to be different, as suggested by the existence of galaxies with late-type morphologies but lower SFRs and redder optical colours that would place them, apparently, on the red sequence \citep{Vandenbergh_1976, Bamford_2009, Wolf_2009}. In contrast, other works have also found some evidence of a gradual decline of the star formation in star-forming objects \citep{Poggianti_2008,Vulcani_2010,Paccagnella_2016} towards denser regions in clusters, so the situation is still somewhat unclear.

Exploring the internal and external processes influencing galaxy evolution requires detailed observations of a large sample of galaxies with a broad range of properties residing in different environments.
Aiming at providing new clues on our understanding of these processes, 
the OMEGA survey (OSIRIS Mapping of Emission-line Galaxies in A901/2) was designed to study the star formation properties and AGN activity of the galaxies in the evolving multi-cluster system A901/2 at $z\sim0.167$. Based on tuneable filter observations with the OSIRIS instrument at the GTC, the survey maps the \Ha($\lambda6563$\AA) and \NII($\lambda6548$\AA{}, $\lambda6584$\AA) emission lines for a large sample of galaxies in an area of the sky of $\sim0.51 \times 0.42$~deg$^2$, equivalent to $\sim5.2 \times 4.3$~Mpc$^2$ at $z\sim0.167$. Our observations are complemented by a wide range of multi-wavelength data already available for A901/2 (XMM-Newton, HST imaging, Spitzer, GALEX, COMBO-17), that has also been subject of study by the STAGES collaboration \citep{Gray_2009}. 

In \citet[][hereafter Paper~I]{chies_2015} we presented a detailed description of the survey, the design of the observations and the results obtained from the study of the two densest regions of A901/2. In this paper, we extend our analysis to a much larger area covering the entire structure, and explore star formation and AGN activity in a significantly larger sample of galaxies distributed over a wider range of environments.
 
This Paper~is structured as follows: in Section \ref{sec:stagesdata} we describe the data products available from previous work that are used here; in Section \ref{sec:datareduction} we explain the methods employed for the reduction and analysis of the data; Section \ref{sec:sampledefinition} contains the criteria chosen to select our galaxy samples; in Section \ref{sec:whandiag} we explore the properties of AGN galaxies; in Section \ref{sec:int_SF} we analyze the integrated star formation properties of the cluster galaxies and in Section \ref{sec:summary} we present a summary and the conclusions of our work. Throughout this paper we adopt an $H_0=70\;\mathrm{km}\,\mathrm{s}^{-1}\,\mathrm{Mpc}^{-1}$, $\Omega_\mathrm{m}=0.3$ and $\Omega_{\Lambda}=0.7$ cosmology. 
 
\section{Available data for the A901/2 system}
\label{sec:stagesdata}
Our study benefits from the many data products already available from a number of previous studies of the A901/2 multi-cluster system. Photometric redshifts, stellar masses, visual morphologies, and spectral-energy-distribution (SED) types of the galaxies were obtained within the STAGES collaboration \citep{Gray_2009} and are described in Paper~I. We note here that the mass estimates of the most massive spirals (log$M_\star$/M$_{\odot}$ $>$ 11) suffer from aperture effects \citep{Gray_2009} and therefore these objects are not considered when galaxy mass is used in our analyses. Based on their SED type, galaxies are divided into normal star-forming galaxies or \emph{blue cloud}, passive ones or \emph{old red} and dust-reddened star-forming galaxies or \emph{dusty reds}. The term \emph{dusty red} was introduced by \citet{Wolf_2005} and it refers to galaxies that show red colours but still host star formation typically $\sim4$ times lower than that in blue spirals at fixed stellar mass \citep{Wolf_2009}. The term \emph{dusty} is perhaps misleading because these objects do not have more dust than normal star-forming galaxies. However, because their star formation is suppressed, their dust content appears relatively high compared with their star formation. Perhaps, a better name would be \emph{red star-forming}, but we keep the term \emph{dusty red} throughout the paper for consistency with previous published works. 

We also benefit from the availability of spectroscopic redshifts for 356 bright cluster galaxies obtained with the AAT 2dF spectrograph and for 182 disk galaxies obtained with the VIMOS instrument at the VLT \citep{Boesch_2013, Boesch_2013b}. Moreover, we have low-resolution prism spectroscopy from the PRIsm MUlti-object Survey \citep[PRIMUS,][]{Coil_2011}. We refer the reader to Weinzirl et al. (in prep.) for a thorough explanation of the redshift surveys and a study of the properties of galaxies with spectroscopic redshifts.  


\section{OSIRIS data reduction and analysis}
\label{sec:datareduction}
In Paper~I we described the methods adopted for the reduction and analysis of the OSIRIS data from the two densest regions. These methods were accurate enough to probe the potential of our observations, one of the primary goals of that paper. However, further work with the data has led to the development of significant improvements in the sky subtraction and wavelength calibration that increase the quality of our data products and consequently the accuracy of our measurements. These modifications are based on the methods described in \citet{Weinzirl_2015}. For the benefit of the reader, we briefly describe here the data reduction and analysis processes. More detailed descriptions can be found in Paper~I and \citet{Weinzirl_2015}. 

The first steps of the data reduction (bias subtraction, overscan removal and flat-fielding) were performed using the OSIRIS Offline Pipeline software \citep[OOPs,][]{OOPs}. This pipeline was also used in Paper~I to subtract the sky emission, which consists of a series of sky rings around the optical centre that are a consequence of the wavelength radial dependence. However, further work with the data showed that the sky subtraction performed by OOPs introduces a significant number of negative pixels around the galaxies, a pattern that is more apparent when the images are stacked. This is due to the relatively small ``dithering'' offsets used by OOPs when creating the median sky image. These negative values for individual pixels are of the order of $\lesssim$5$\%$ of the total flux. Given the strong radial dependence of the sky background, larger XY offsets cannot be used. A new sky subtraction method that takes advantage of the slow azimuthal variation of the sky background was developed by \cite{Weinzirl_2015}. Using this method, the sky subtraction was significantly improved, resulting in an increase in the mean emission-line fluxes of $\sim40\%$. See \citet{Weinzirl_2015} for details. 

After subtracting the sky, the best astrometric solutions for the images are found using the IRAF routines CCMAP and WCSCTRAN. Since the observations were taken during nights with different atmospheric seeing conditions, the images of all observed fields are convolved with an appropriate Gaussian function to match the worst seeing ($\sim$ 1.2 arcsec, 5 pixels). The flux calibration is performed using a set of stars (between 15 and 20 per image) that are used to generate a wavelength-dependent zero-point function for each image, in the way described in Section 3.3 of Paper~I. 

In Paper~I the wavelength calibration for two of the fields (numbers 21 and 22) was carried out using the solution proposed by \cite{Gonzalez_2014}, which is aimed at achieving accuracies of 1--2\AA. The precision of this crucial calibration could only be tested with our data if an object is observed in pointings centred at different positions. Unfortunately, fields 21 and 22 do not overlap with each other (see Figure~1 in Paper~I), thus preventing us from testing the accuracy of the wavelength calibration at the time. However, the data from the 20 observed fields contain a large number of galaxies that were observed in two or even three pointings (see Figure~1 in Paper~I). When comparing the spectra of these galaxies, we find offsets as high as 7\AA, much larger than expected. These offests are primarily due to inaccuracies in the determination of the central wavelength of each pointing. Although these uncertainties in the wavelength calibration do not affect significantly the estimation of the fluxes in \Ha and \NII, they compromise the calculation of the redshifts of the galaxies and consequently the study of the cluster dynamics, which are also being analysed (Weinzirl et al. in prep). To overcome these difficulties we adopt the improved wavelength calibration developed by \cite{Weinzirl_2015}, which uses the known wavelength of several sky emission lines to obtain wavelength solutions accurate to $\pm1$\AA{} for our data.

The next step in the data analysis is the measurement of the aperture photometry for all the galaxies in each image. As mentioned above, one of the main goals of the OMEGA survey is to study star formation and AGN activity in the cluster galaxies. Since the radiation from an AGN host comes mainly from the nuclear parts, we probe the presence of AGN by measuring the galaxy flux within a PSF-matched aperture of radius $1.2\;\mathrm{arcsec}$ ($R_{\rm PSF}$). Integrated fluxes, necessary to measure total SFRs, are obtained using a larger aperture that encompasses the full extent of the galaxies ($R_{\rm tot}$). Thus, for every galaxy we produce one integrated and one nuclear spectrum, which are the same for unresolved objects. More details on the aperture photometry procedure can be found in Section 3.4 in Paper~I.  

The last steps of the analysis are building the spectra from the photometric measurements and estimating the line fluxes and positions, and the continuum levels. These steps are described in detail in Sections 3 and 4 of Paper~I. The line fluxes (\FHa, \FNII), equivalent widths (\WHa) and redshift of the galaxies ($z$) that we use throughout this paper correspond to the median values of their probability density distributions obtained from Markov chain Monte Carlo (MCMC) spectral fitting techniques. These probability density distributions are also  used to estimate the errors associated with these parameters. We refer the reader to section 4.1 of Paper~I for a thorough explanation of the spectral fitting process.

\subsection{Correcting for underlying stellar absorption}

The emission lines in our spectra originate from the gas ionised by either hot stars or AGN. However, the stellar populations inhabiting these galaxies also leave an imprint on the galaxies' spectra by absorbing part of the light in their stellar atmospheres. As a consequence of this absorption, the total flux in emission might be underestimated. We therefore need to account for the absorption from the underlying stellar population in order to obtain an estimate of the true total \Ha fluxes and SFRs. To do that, we use the PEGASE models \citep{Fioc_Rocca_1997} that were employed by \citet{Wolf_2005} to fit the COMBO-17 SEDs. These models give an estimate of the expected absorption in \Ha for a given age of the SED templates. We use the \Ha absorption of the model corresponding to the age of the best-fitting SED template (available in the main STAGES catalogue of \citealt{Gray_2009}) to correct the \Ha fluxes and equivalent widths from the stellar absorption. 

In young objects ($\sim$$0.5$--$1\,$Gyr), the expected absorption is typically $\sim6$\AA, whereas in old objects ($\ge3\,$Gyr) the absorption becomes less than $2$\AA. Because less massive galaxies tend to host younger stellar populations, the average expected \Ha absorption correlates with galaxy stellar mass, changing from $\sim5.5$\AA\ at $10^9$M$_\odot$ to $\sim4$\AA\ at $10^{10}$M$_\odot$ and $\sim2$\AA\ at $10^{11}$M$_\odot$. All our \Ha estimates and therefore the SFRs are corrected for stellar absorption in this way, and the corresponding uncertainties are propagated when estimating the relevant errors. On average, this correction reduces the SFR of the galaxy sample by $\sim20\%$, being proportionally less important at high galaxy masses. These corrections are only applied to the sources that satisfied the selection criteria defined in Section~\ref{sec:sampledefinition}.

\section{Sample definition}
\label{sec:sampledefinition}

Our parent sample is drawn from the STAGES catalogue, where we select all the objects with $R$-band magnitude $(m_R)\leq23$ that are detected in both COMBO-17 and the STAGES HST imaging (\hbox{{\tt combo\_flag}$\ >1$} and \hbox{{\tt stages\_flag}$\ > 1$} in \citealt{Gray_2009}). The objects are also required to lie within the footprint of the 20 observed OMEGA fields (Paper I), yielding $\sim7600$ objects. Aperture photometry is carried out on all the OMEGA images at the position of these sources, and spectra are constructed and analysed as described in Section \ref{sec:datareduction}. 

To construct a sample containing only emission-line galaxies from the A901/2 multi-cluster system we investigate the presence of emission lines in the galaxy spectra and assess whether they belong to this system based on the COMBO-17 photometric redshifts. This section describes in some detail the selection process we follow.   

\subsection{Spectral fitting criteria}
\label{Fittingcuts}

The presence of emission lines in the spectra is evaluated through the spectral fitting process. Since the integrated ($R_{\rm tot}$) and nuclear ($R_{\rm PSF}$) spectra can vary significantly in galaxies hosting an AGN and those with non-homogeneous distributions of star formation, in the following we define different samples for each set of spectra (nuclear and integrated).

As explained above, we use MCMC techniques to fit each observed spectrum using a model containing three emission lines (\Ha and the two \NII lines) on top of a continuum. We consider that the \Ha line is detected in emission if the following criteria are fulfilled:

\begin{itemize}
\item $F\mathrm{_{H\alpha} \geq 3 \times 10^{-17}erg^{-1}cm^{-2}s^{-1}}$;
\item $W\mathrm{_{H\alpha} \geq 3.0}$\AA;
\item the \Ha line is inside the observed wavelength range, with its peak more than half the instrumental resolution (i.e., $3.75$\AA) from the edges of this range.
\end{itemize}

These criteria ensure the detection of \Ha, and are used to build the main sample. We note here that the selection criteria have slightly changed from Paper I, following the improved sky subtraction and the application of a more robust and systematic visual inspection to avoid unreliable or spurious detections. Since in some of the scientific applications of these data we also need to analyse the \NII\ line (e.g., when studying AGN activity), we also build a sub-sample of the main sample by requiring that \NII\ is contained in the observed wavelength range. Specifically, for this sub-sample we require that the last criterion is also met by the $[\mathrm{N}\textsc{ii}]\lambda6584$\AA\ line.

\subsection{Photometric redshifts criteria}
\label{photometricredshiftcuts}

Apart from selecting galaxies with clear emission lines in their spectra, it is crucial to ensure that these lines are indeed \Ha and \NII. Although our program is designed to map these lines in galaxies residing in the multi-cluster system, in particular with redshifts in the [0.150, 0.176] interval, we also obtain spectra of objects in front and behind the system for which any detected emission line will not be \Ha or \NII. Fortunately, the COMBO-17 photometric redshifts are accurate enough to remove any possible interlopers from our sample (see Section~\ref{sec:stagesdata}). For instance, contamination from higher-redshift $[\mathrm{O}\textsc{iii}]\lambda5007$\AA\ emission at $z\sim0.53$ or H$\beta$ at $z\sim0.57$ can easily be rejected.

With this in mind, emission-line objects selected applying the criteria described in Section~\ref{Fittingcuts} are only accepted in our sample if their photometric redshifts obey the condition $0.150 - 3\Delta z \le z_{\rm phot} \le 0.176+3\Delta z$, where $\Delta z = 0.008$ is the photometric redshift uncertainty at $m_R = 20.5$, the median apparent magnitude of the emission line galaxy sample \citep[see eq.~5 of][]{Gray_2009}. Based on this, only emission-line galaxies with photometric redshifts in the interval [0.126, 0.200] are accepted in our sample. The available spectroscopic data (cf. Section~\ref{sec:stagesdata}) allows us to check the effectiveness of the photo-$z$ selection. There are 149 emission-line sources in our OMEGA sample having spectra from 2dF/VIMOS.  Of these, 134/149 are true members with $z_{\rm spec}$ in the [0.150, 0.176] interval. The $z_{\rm phot}$ criterion selects 136/149 sources.  Only 4/136 of these are false positives (meeting the $z_{\rm phot}$ criterion but not the $z_{\rm spec}$ one). This means the sample is over 97\% pure. There are two false negatives
(meeting the $z_{\rm spec}$ criterion but not the $z_{\rm phot}$ one), so the sample is more than 98\% complete. Of course, this test is only representative of the galaxies with magnitudes comparable to those in the spectroscopic redshift sample (brighter than $m_R \sim 22$), while the OMEGA emission-line galaxy sample reaches $\sim1\,$mag deeper. We expect the purity and completeness of the full sample to be somewhat worse, but given the quality of the COMBO-17 redshifts, we are confident that this issue will not significantly affect our results. 

The final step in the selection of the sample is a visual inspection of the OSIRIS spectra to ensure that all the selected galaxies exhibit bona-fide emission lines. As discussed in \cite{Sanchez_2015}, this step is necessary to reject spurious detections which are due to instrumental and observational artefacts that cannot be efficiently detected automatically. Three co-authors (AA-S, BRdP and TW) carried out this inspection, and galaxies were accepted or rejected according to a majority vote.

Applying the selection criteria explained in this section together with those based on the spectral fitting
(Section~\ref{Fittingcuts}) our sample contains a total of 439 galaxies with \Ha\ detected in the total aperture spectra, of which 321 also have \NII\ in the observed wavelength range. When considering the $R_{\rm PSF}$ aperture spectra (used here to study AGN activity) the sample contains 360 galaxies with \Ha\ detected and \NII\ in the observed wavelength range.

\subsection[Detection efficiency and limits]{Detection efficiency and limits}
\label{completeness}

\begin{figure}
\begin{center}
\includegraphics[width=0.49\textwidth]{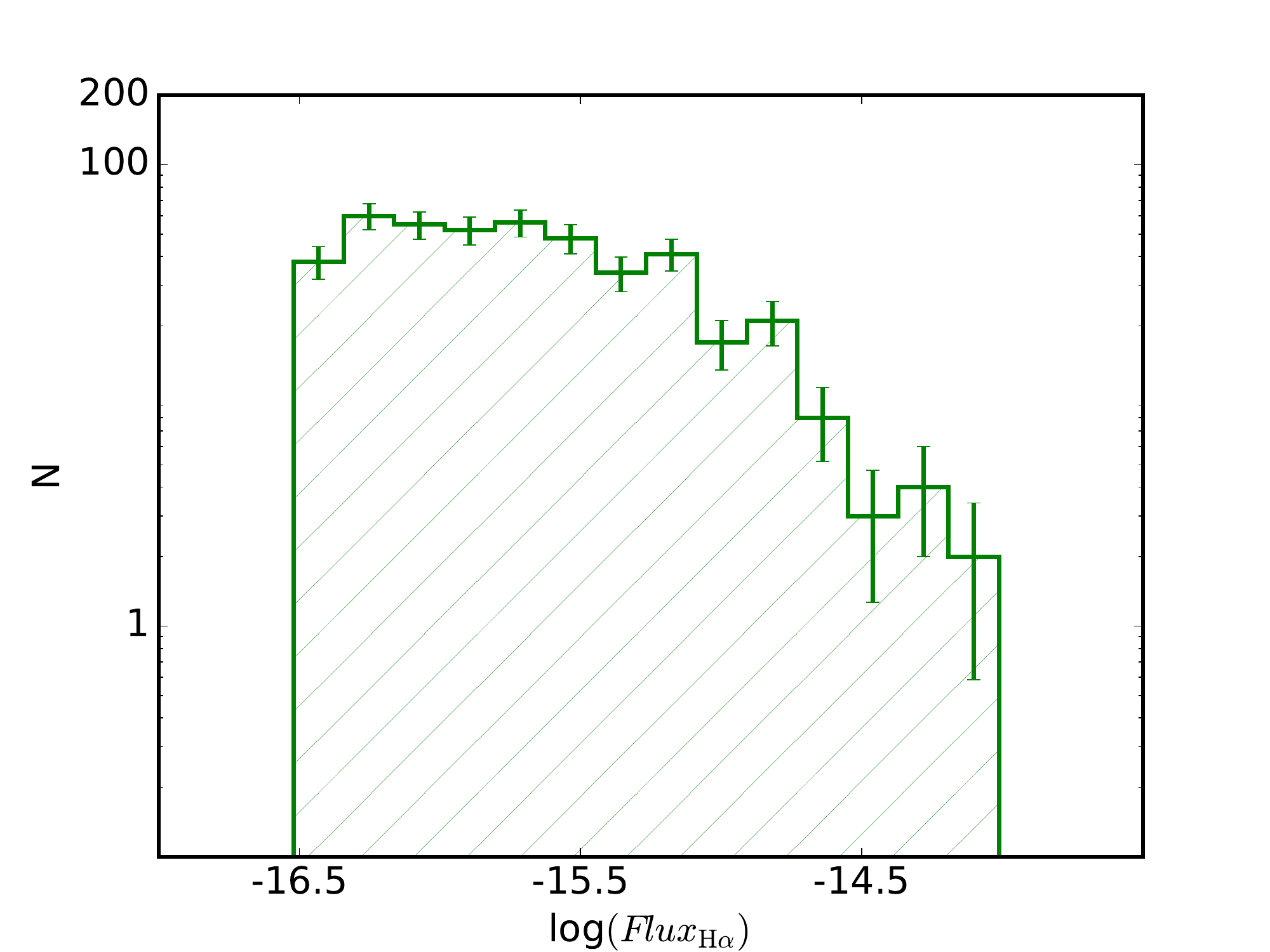}
\caption[\Ha flux distribution for the emission-line galaxies]{The distribution of the \Ha flux for the 439 ELGs in our sample, where the error bars are $1/\sqrt{N}$.}
\label{FHa}
\end{center}
\end{figure}

\begin{figure}
\begin{center}
\includegraphics[width=0.49\textwidth]{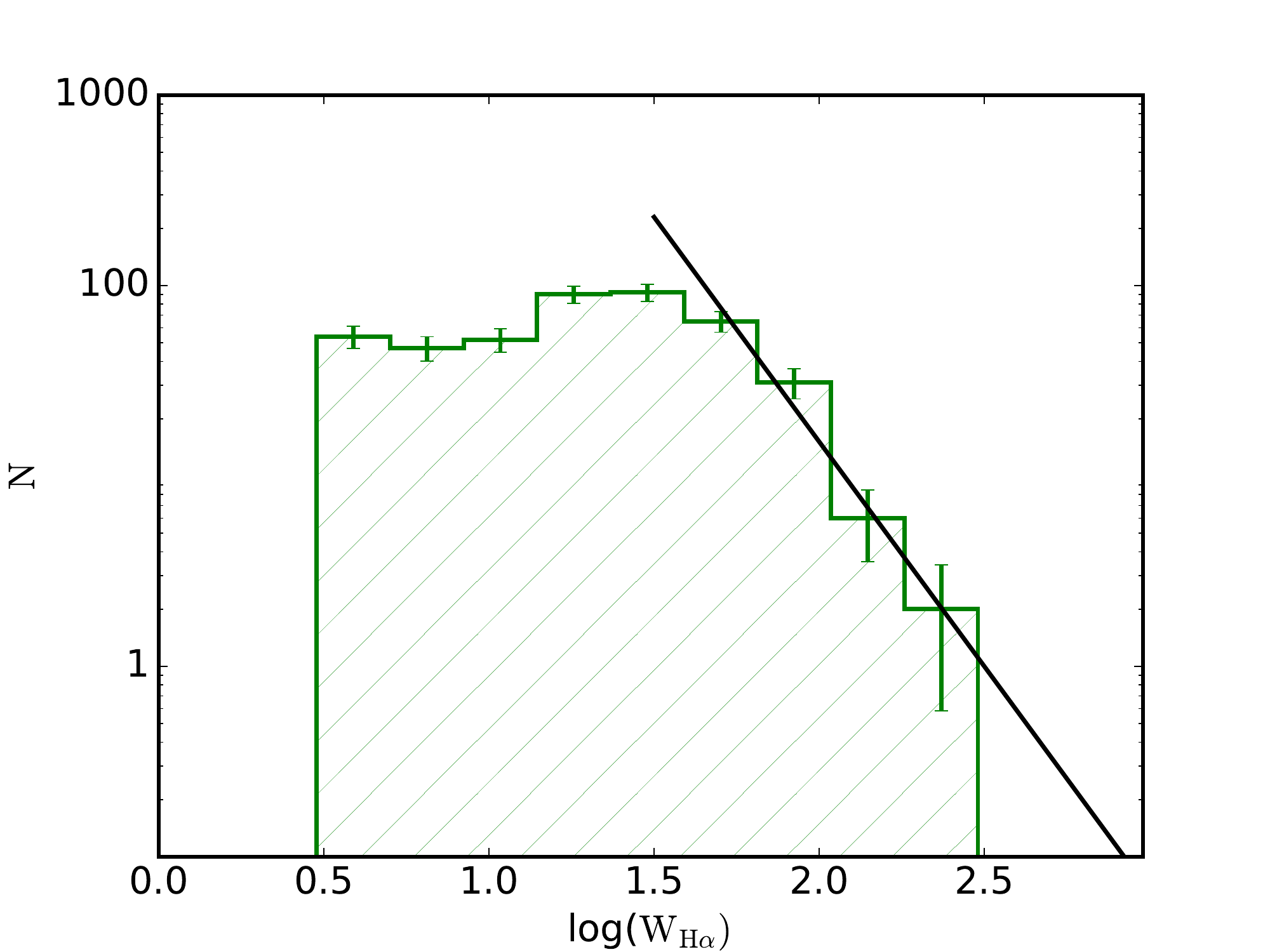}
\caption[\WHa flux distribution for the emission-line galaxies]{The distribution of \WHa for the 439 ELGs in our sample, where the error bars are $1/\sqrt{N}$. For equivalent widths larger than $\sim30$\AA\ the distribution declines steeply following a power low with slope $-2.3$ and intercept $5.9$ (solid line).}
\label{WHa}
\end{center}
\end{figure}

The breadth and depth of our study is constrained by our ability to reliably detect the \Ha line. As described in Paper~I, the probability of detecting \Ha in emission depends on its total flux (\FHa) and equivalent width (\WHa). The distributions of \FHa and \WHa for the 439 \Ha-detected galaxies are plotted in Figures \ref{FHa} and \ref{WHa}, respectively.

We are able to detect galaxies with \Ha fluxes as low as $3 \times 10^{-17}\,$erg$\,$cm$^{-2}\,$s$^{-1}$, and as high as $5 \times 10^{-15}\,$erg$\,$cm$^{-2}\,$s$^{-1}$. We sample higher \Ha fluxes than in Paper~I as a consequence of the improved sky subtraction that we apply here (see Section \ref{sec:datareduction}). The \WHa distribution is broadly compatible with the one found by \cite{Balogh_2004_ecology}. For equivalent widths larger than $\sim30$\AA, the distribution declines steeply, following a
power law with slope $-2.3$, and reaching as high as $\sim300$\AA. At lower equivalent widths, the distribution flattens out, reaching $\sim3$\AA. Incompleteness probably kicks in below $\sim5$\AA. When interpreting our data it is important to bear in mind that the sensitivity of our survey declines for lower equivalent widths. We discuss the effect of our flux and equivalent-width detection limits later in the paper.

The \Ha\ detection process is complex and it would be very difficult to model accurately the detection limits and the completeness of our sample from first principles. For this reason, in Paper~I we devised an empirical procedure to test our ability to detect \Ha\ emission and quantify the completeness of the OMEGA survey. This procedure takes advantage of the wealth of data available from the STAGES data base, and is followed here for the whole sample. A detailed description can be found in Paper~I, but we repeat the main points here for completeness. 

Two types of galaxies are expected to host star formation and therefore exhibit \Ha emission. The first type are blue star-forming galaxies in the \emph{blue cloud}. The second, perhaps less obvious type, are the so-called \emph{dusty red} spirals which, despite having optical colours that would apparently place them on the optical red sequence, still have a certain degree of star formation \citep{Wolf_2009}. We will compare our \Ha detections with the numbers of
blue-cloud and red spiral galaxies (all of them star-forming candidates, and thus expected \Ha emitters) in the STAGES
catalogue in order to evaluate how efficient the OMEGA survey is at detecting star-forming galaxies and thus quantify the sample completeness.

Using the STAGES photo-$z$'s and SED classes (Section \ref{sec:stagesdata}) we find that in the parent sample (Section~\ref{sec:sampledefinition}) there are $\sim 7600$ objects, of which $1244$ are classified as cluster members (Section \ref{photometricredshiftcuts}). Of these, 316 are \emph{dusty reds} and 630 belong to the \emph{blue cloud}. We securely detect \Ha in 69 \emph{dusty reds} and in 330 \emph{blue cloud} galaxies, yielding preliminary \Ha-detection efficiencies of  $\sim22$\% and $\sim52$\% for each galaxy class respectively. 
However, these numbers assume that the STAGES star-forming galaxy samples are complete and uncontaminated, which they are not. Furthermore, we need to quantify how the detection efficiency depends on galaxy mass/brightness. 

As described in \citet{Gray_2009}, the cluster membership was determined using COMBO-17 photometric redshifts. Despite the relatively high accuracy of these redshifts, the cluster sample suffers from a certain degree on incompleteness and contamination, particularly at faint magnitudes. Therefore, the ``true'' (corrected for incompleteness and contamination) number of cluster galaxies 
$N_\mathrm{c17}^\mathrm{c}$ can be estimated as
\begin{equation}
\label{corr1}
N_{\rm c17}^{\rm c}=\frac{N_{\rm cmc17}(1-\mathit{cont}_\mathrm{frac})}{\mathit{comp}_\mathrm{frac}},
\end{equation}
where $N_{\rm cmc17}$ is the observed number of cluster members based on their photometric redshifts (see Section \ref{photometricredshiftcuts}). In this equation, $\mathit{comp}_\mathrm{frac}$ and $\mathit{cont}_\mathrm{frac}$ are the magnitude-dependent completeness and contamination fractions respectively (see Fig.~14 of \citealt{Gray_2009}). A similar calculation is done to estimate the 
`true' number of \emph{blue cloud} (BC) cluster galaxies and \emph{dusty red} (DR) cluster galaxies (cf. Paper~I). 

\begin{figure}
\begin{center}
\includegraphics[width=0.5\textwidth]{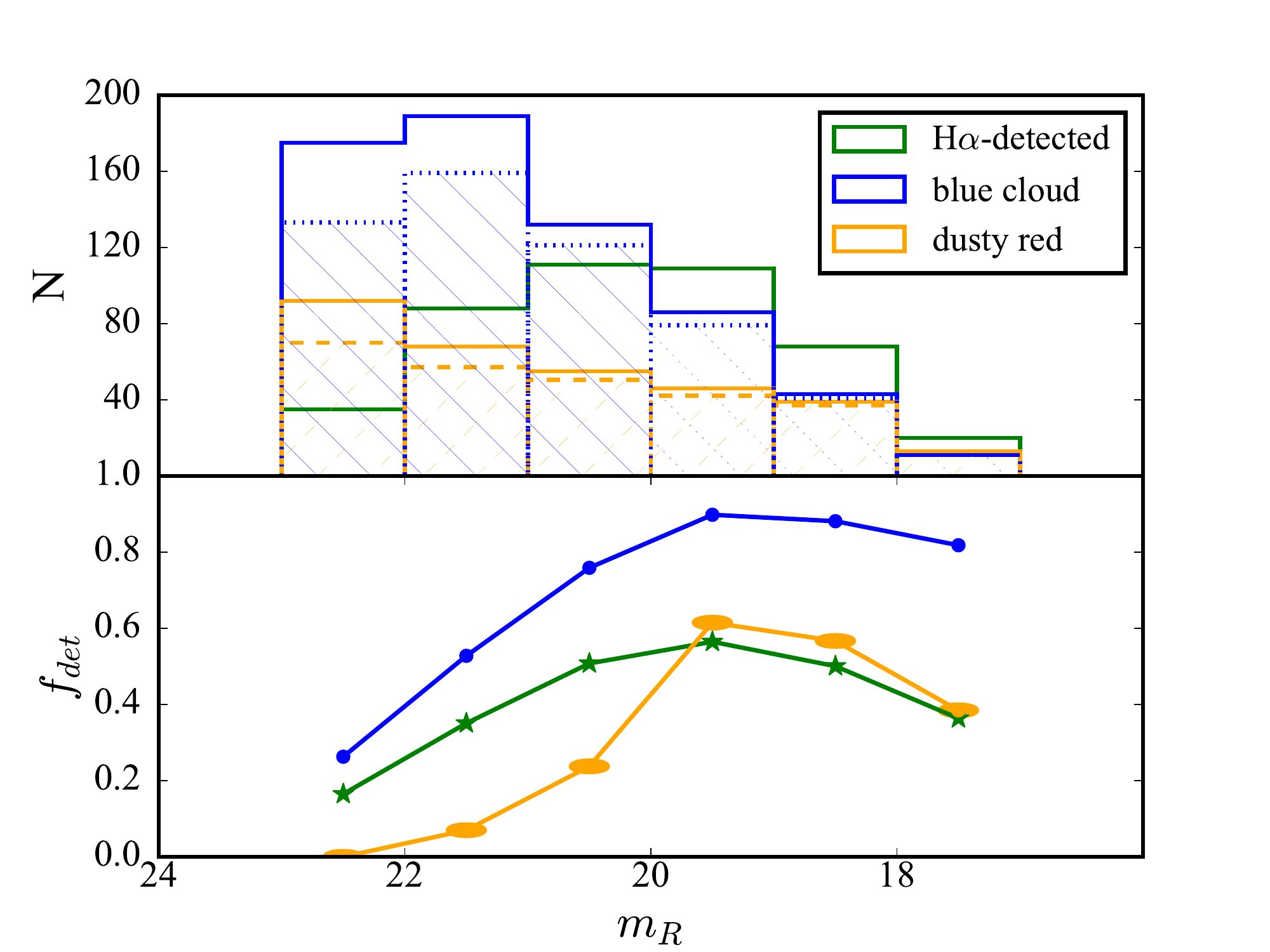}
\caption[OMEGA detection limits]{\textit{Top panel}: $m_R$ distribution of the \Ha-detected galaxies, \emph{blue cloud} and \emph{dusty red} galaxies determined to be cluster members using the COMBO-17 photometric redshifts (see Section \ref{photometricredshiftcuts}). The dashed histograms correspond to \emph{blue cloud} and \emph{dusty red}  samples corrected for completeness and contamination as discussed in the text.
\textit{Bottom panel}: The \Ha detection rate, expressed as the fraction of \Ha-detected cluster members (green stars), \emph{dusty reds} (yellow ellipses) and \emph{blue cloud} galaxies (blue circles), as a function of $m_R$. These fractions take into consideration the completeness and contamination correction estimated using Equation~\ref{corr1}. }
\label{Rmag}
\end{center}
\end{figure}

In the top panel of Figure~\ref{Rmag} we show the $m_R$ distribution of the \Ha-detected galaxies, \emph{blue cloud} and \emph{dusty red} cluster galaxies. The dashed histograms correspond to the last two samples corrected for completeness and contamination. The fraction of cluster galaxies (total, \emph{blue cloud} and \emph{dusty red}) that are detected in \Ha are shown in the bottom panel of the same figure (green stars, blue circles and orange ellipses, respectively). This fraction takes into account the completeness and contamination correction estimated using Equation~\ref{corr1}.

Here we can see a clear difference in our efficiency of detecting star formation in \emph{blue cloud} and \emph{dusty reds}. Whereas we effectively detect \Ha for all \emph{blue cloud} cluster members brighter than $m_R \sim 20$, the fraction of \emph{dusty reds} detected is always lower. This is not surprising since the star formation of \emph{dusty reds} is suppressed by a factor of 4 on average compared with \emph{blue cloud} galaxies of similar stellar mass \citep{Wolf_2009}. This suppression of star formation implies reductions in both \Ha flux and equivalent width. For both types of galaxies, the completeness drops off for fainter magnitudes, as expected. This drop is more pronounced for \emph{dusty reds} fainter than $m_R \sim 19.5$, where \WHa declines dramatically. In \emph{blue cloud} galaxies the detection rate drop with magnitude is slower because when we move towards fainter and less massive galaxies, the reduction in SFR (or \Ha flux)  is partially compensated by an increase in specific SFR and thus \WHa.   

Note that the apparent drop in \Ha detection efficiency for galaxies brighter than $m_R \sim 19$ is somewhat uncertain due to low-number statistics in the brightest bins. However, more massive galaxies tend to have lower specific star formation rates and thus low \WHa, making them harder to detect. The detection rate for all cluster members as a function of $m_R$ is relatively flat since at lower masses (fainter magnitudes) a greater fraction of cluster galaxies are star forming, while the fraction of star-forming galaxies declines at higher masses.  

\begin{figure}
\begin{center}
\includegraphics[width=0.49\textwidth]{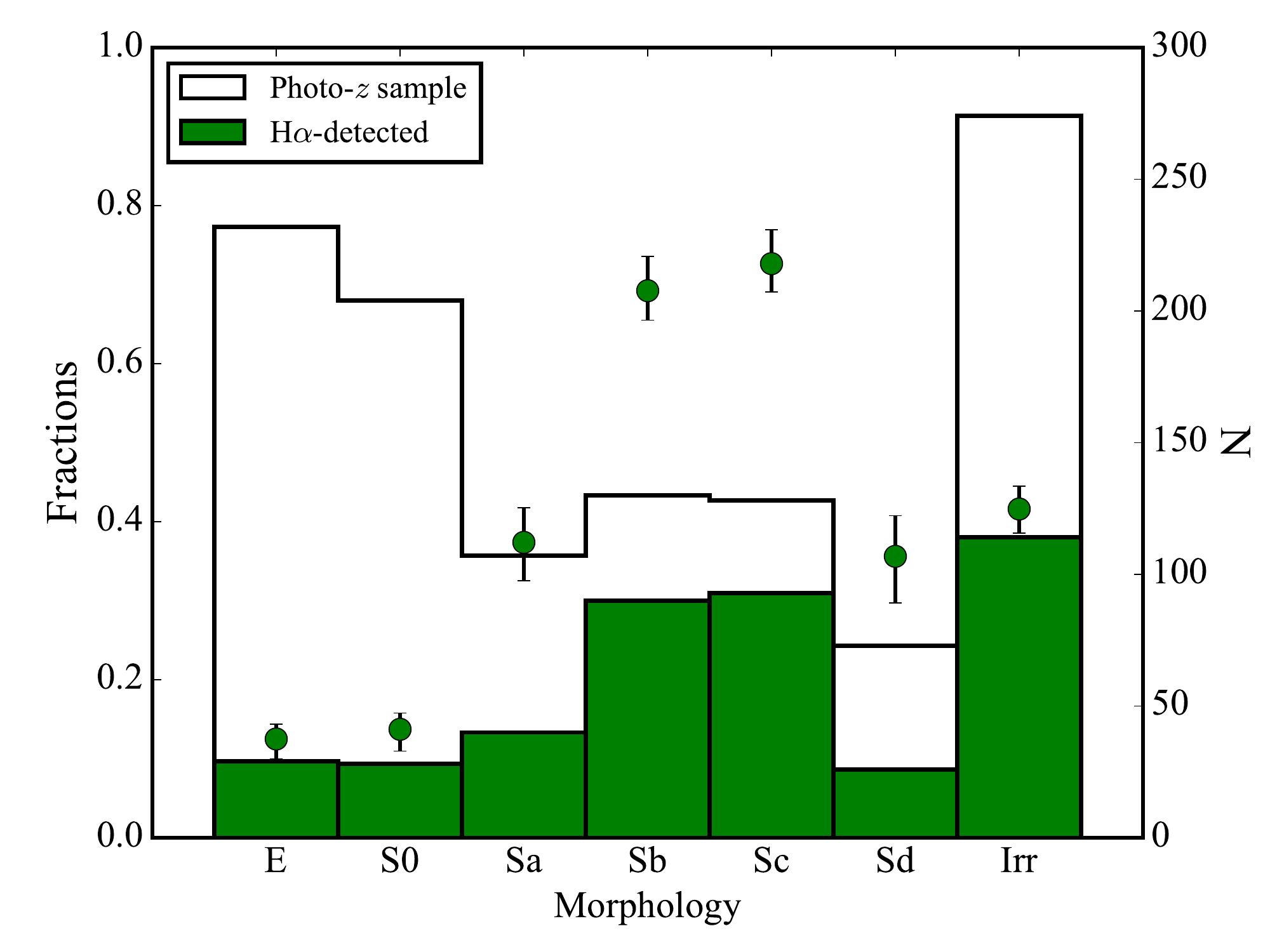}
\caption[morphological fractions]{Fraction (filled circles and left scale) and total number of galaxies (histograms and right scale) detected as \Ha emitters with respect to the parent galaxy population using the same stellar mass and redshift selection criteria (``Photo-$z$ sample'', see Section~\ref{photometricredshiftcuts}). As expected, irregular and spiral galaxies, specially Sb and Sc types, have much higher \Ha detection rates than ellipticals and S0s. }
\label{morphofraction}
\end{center}
\end{figure}

As a complement to the study of the global detection efficiency, we also explore which morphological types are more frequently detected as H$\alpha$ emitters in OMEGA. In Figure \ref{morphofraction} we show, for each morphological type, the fraction of galaxies detected in H$\alpha$ with respect to the parent galaxy population with the same stellar mass and photometric redshift selection criteria (``Photo-$z$ sample'', Section \ref{photometricredshiftcuts}). As expected, spiral and irregular galaxies are the morphological types most frequently detected as \Ha emitters, whereas we only detect \Ha in $\sim10\%$ of ellipticals and S0s. The drop in detection rate for Sd and Irregular galaxies is explained by the fact that these galaxies tend to be less massive and fainter, and, for a given specific SFR, their SFRs and \Ha luminosities are lower.

\subsection{Environment Definition}
\label{envdef}

The data presented in Paper~I focused only on the two densest regions of the multi-cluster system, effectively sampling one single environment. Here we cover a much larger field and therefore a broader range of densities. Since one of our main interests is to study the influence of environment on the properties of galaxies, we need to explicitly define how we measure environment. Previous works using the STAGES data, such as \citet{Wolf_2009} and \citet{Maltby_2010}, have defined their field and cluster samples using the photometric redshift distribution. OMEGA is designed to only target galaxies that are in the multi-cluster system, avoiding field galaxies, but we  need a measure to account for the different environments in which these galaxies reside \emph{within} the system.

It has been previously shown that, in the A901/2 system, galaxy property trends with environment are more sensitive to local galaxy density than to cluster-centric distance \citep{Lane_2007}. Therefore, here we evaluate the environment using the stellar mass surface density which, unlike galaxy number surface density, does not depend strongly on the selection of the underlying sample (either by luminosity or by stellar mass; \citealt{Wolf_2009}). Following the procedure of \citet{Wolf_2009} we use as our density measure the parameter $\Sigma^{\rm M}_{300kpc} (>10^{9}$M$_{\odot}$), in units of M$_{\odot}$/Mpc$^{2}$. This parameter measures the stellar mass surface density inside a fixed aperture of radius $r$~=~300kpc at the redshift of the multi-cluster system ($z = 0.167$), including only those galaxies with log $M_{\star}/$M$_{\odot}$ $>$ 9.  Following \citet{Maltby_2010}, galaxies with $\Sigma^{\rm M >10^{9}{\rm M}_{\odot}}_{300kpc}$~$>$~12.5 {\rm M}$_{\odot}$Mpc$^{-2}$ will be considered to belong to the cluster \emph{cores} whereas the rest of cluster galaxies will be associated with the \emph{infall} regions of the system. Stellar mass surface densities are available for all the objects that are classified as cluster members by \citet{Gray_2009}, but, since our criteria to select OMEGA emission-line galaxies as cluster members is slightly different (cf. Section~\ref{photometricredshiftcuts}), there are a few galaxies without stellar mass surface density estimates. In total, 381 out of the 439 \Ha-detected galaxies and 306 out of 360 galaxies where \Ha and \NII are detected have stellar mass densities.

\section{The WHAN diagnostics diagram and AGN activity}
\label{sec:whandiag}

The identification of AGN activity using optical emission lines is based on the fact that an accreting black hole produces a harder (more energetic) radiation field than star formation. As a consequence of their different energetic input, the atomic energy levels ionized and the subsequent intensities of the emission lines after recombination are different, making the ionization source traceable. Historically, the most-widely used emission-line diagnostic is based on the BPT diagrams (\citealt{BPT_1981}), of which the most common one uses four strong emission lines [O \textsc{iii}], H$\beta$, \NII and \Ha. However, in order to obtain reliable measurements of these lines one needs access to different parts of the spectrum with relatively high signal-to-noise.

\begin{figure}
\begin{center}
\includegraphics[width=0.49\textwidth]{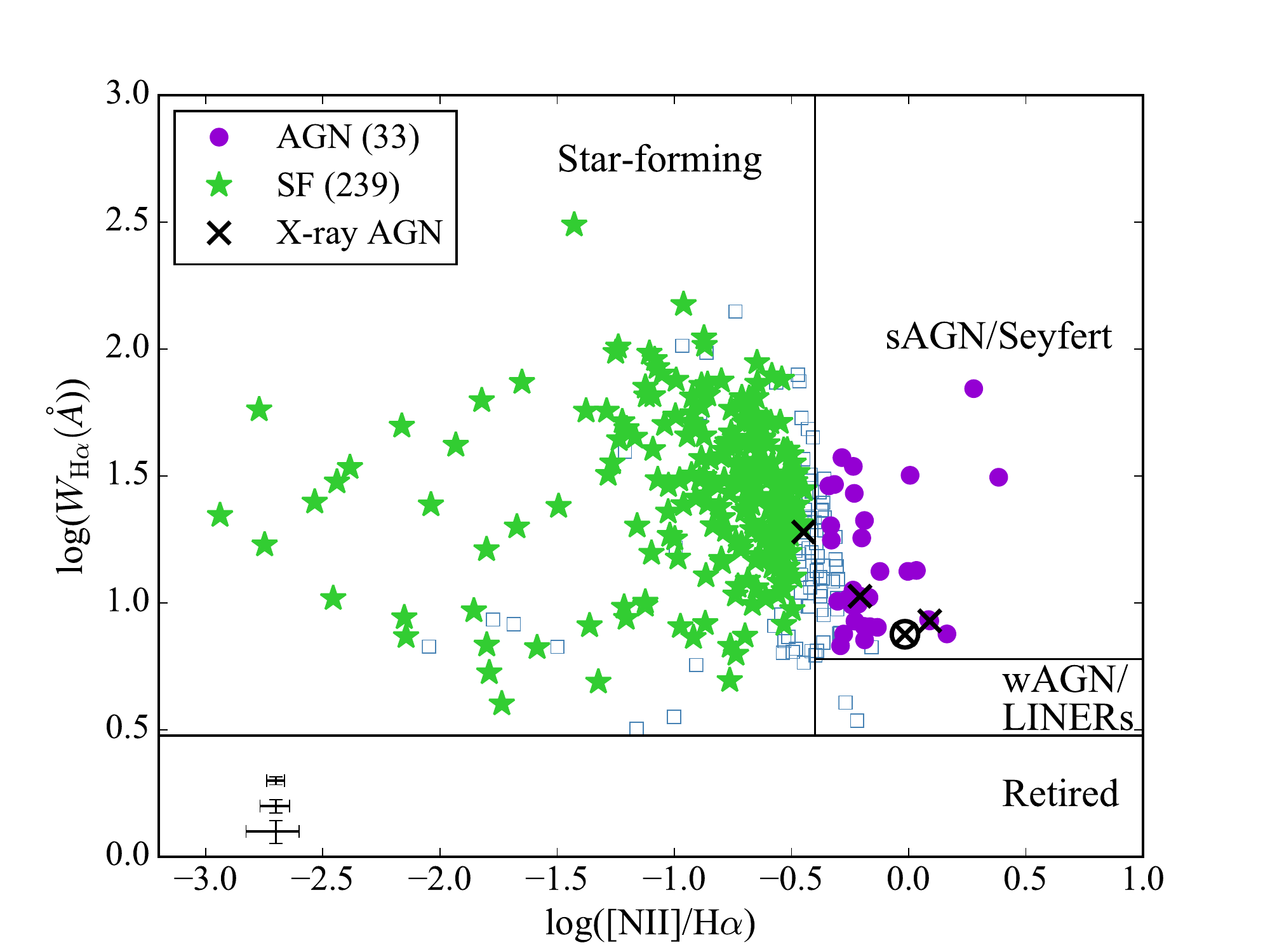}
\caption[WHAN diagram]{The WHAN diagram diagnostic plot (\citealt{Cid_2011}). The vertical and horizontal lines separate regions with different sources of ionization. We only show detections using $R_{PSF}$ aperture measurements where both \Ha and \NII fall in the OMEGA probed wavelength range: a total of 360 ELGs. From these we find 33 secure ($\ge3\sigma$ confidence) AGN hosts (Seyferts; purple circles) and 237 star-forming galaxies (SF; green stars). The blue squares represent galaxies that contain mixed sources of ionisation and/or the classification is more uncertain. The black crosses correspond to the X-ray AGN discussed in Section~\ref{AGNXray} whose OMEGA spectra are shown in Figure~\ref{fig:AGN_spectra} (see text for details). In the bottom-left corner we show the 25, 50 and 75 percentiles of the errors for reference. }
\label{WHAN}
\end{center}
\end{figure}

A more economical way of identifying the sources of ionization in a galaxy using only two emission lines, \Ha and \NII, was proposed by \citet{Cid_2010}. The proposed diagnostic diagram, known as the WHAN diagram, compares the equivalent width of \Ha (\WHa) with the ratio between the fluxes of \NII and \Ha (\ratio). This diagram is designed to detect lower levels of ionization. It is worth mentioning that in nature there is not such a sharp division between systems dominated by star formation and those dominated by AGN activity. In fact, the position of the vertical boundary varies depending on the work used as a reference \citep{Kewley_2001,Kauffmann_2003_agn,Stasinska_2006}. In our case, we employ the position proposed by \citet{Stasinska_2006}, \ratio $= 0.4$, which despite being based on photoionization models, it is the one that best follows the distribution of the SF population in the diagram \citep{Cid_2010}. Furthermore, there is now a growing body of evidence indicating that in some cases, high \ratio ratios can also be caused by shocks \citep{Ho_2014} or other sources of ionising radiation not directly caused by AGN activity \citep{Singh_2013, Belfiore_2016}, although this tends to occur over extended regions of the galaxies rather than the nucleus. In these cases, the extended emission would be more easily detected using a large aperture for the flux estimation (encompassing more regions with high $[{\rm NII}]$/\Ha). However, a larger aperture would reduce the $[{\rm NII}]$/\Ha ratio in the case of AGN-driven nuclear emission, making bona-fide optical AGN more difficult to identify. In practice, given the limited spatial resolution of our ground-based OMEGA images, the statistical properties of our AGN sample change very little when the analysis is carried out with the larger aperture rather than with the smaller one (see below).

Bearing these subtleties in mind, four different populations can be identified in the diagram: \emph{star-forming} galaxies, strong AGN (sAGN) or \emph{Seyferts}, weak AGN (wAGN) or \emph{LINERs}, and the so-called ``retired'' galaxy population (\citealt{Cid_2011}, \citealt{Yan_Blanton_2012}, \citealt{Stasinska_2015}). LINER stands for `low ionisation nuclear emission regions' (as introduced by \citealt{Heckman_1980}).  Therefore, LINERs are galaxies with `LINER-like spectra'. The weak emission lines in these objects can be seen in the nucleus, due to the presence of an AGN, but they may also appear in the galactic disk \citep[e.g.,][]{Belfiore_2016}, where they can be produced by fast shocks \citep{Dopita_1995}, starburst-driven winds \citep{Armus_1990} or diffuse ionized plasma \citep{Collins_2001}. Finally, the ``retired'' galaxies are objects that are no longer forming stars and their gas is believed to be ionized by hot, low-mass stars in the post-AGB phase. They do not have a detectable contribution from an AGN and  their \Ha emission is often extended \citep{Singh_2013, Belfiore_2015,Belfiore_2016}.

The WHAN diagram for the 360 sources whose central spectra cover both \Ha and \NII is shown in Figure~\ref{WHAN}. 
Here we identify which sources are more likely to be hosting an AGN or to be dominated by star formation
using the probability density distributions of \WHa and \ratio obtained from the MCMC fitting of the spectra (cf. Paper I). The sources with probabilities higher than $99.7\%$ ($\simeq3\sigma$ confidence) of being AGN or star-forming are plotted as purple circles and green stars, respectively. Based on this procedure we find 33 objects with high likelihood of being AGN hosts and 239 likely dominated by star formation. For the remaining 90 galaxies (plotted as blue squares) the classification is more uncertain (less than $99.7\%$ probability) or intermediate. 

Using the total flux measurements (within $R_{\rm tot}$) instead of the central ones in the WHAN diagrams yields marginally higher AGN fractions (by $\sim$10\%, or 3 additional sources), implying that only a small fraction of the galaxies classified as AGN might be associated with extended shocks. Unfortunately, the limited spatial resolution of our ground-based data prevents us from carrying out a detailed analysis of the emission outside the region affected by the PSF.

\subsection{OMEGA and X-Ray AGN cross-matching}
\label{AGNXray}

The presence of nuclear activity in galaxies can be observed at different wavelengths. A previous study of AGN activity in the A901/2 multi-cluster system was performed by \cite{Gilmour_2007} using \textit{XMM}-Newton X-ray data. Since this type of data are able to detect luminous X-ray AGN but do not detect lower levels of nuclear activity and/or Compton-thick objects, it can be used as a complement to the optical AGN detected in OMEGA in order to have a wider coverage of AGN luminosities.
It is therefore interesting to compare these two methods of detecting AGN activity to explore their efficiency and whether some sources are detected in both wavelength ranges.

\begin{figure}
\begin{center}
\includegraphics[width=0.36\textwidth]{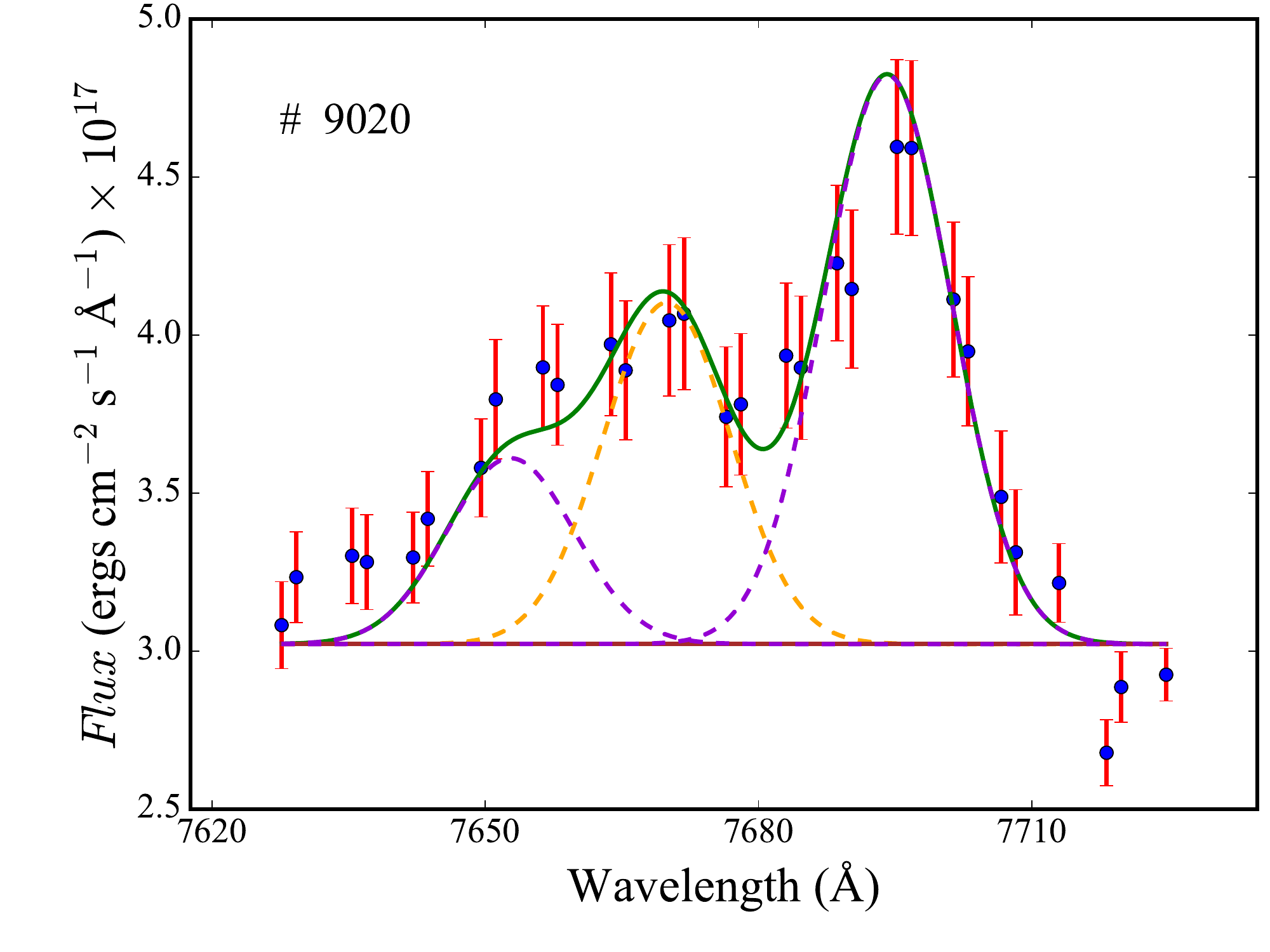}
\includegraphics[width=0.36\textwidth]{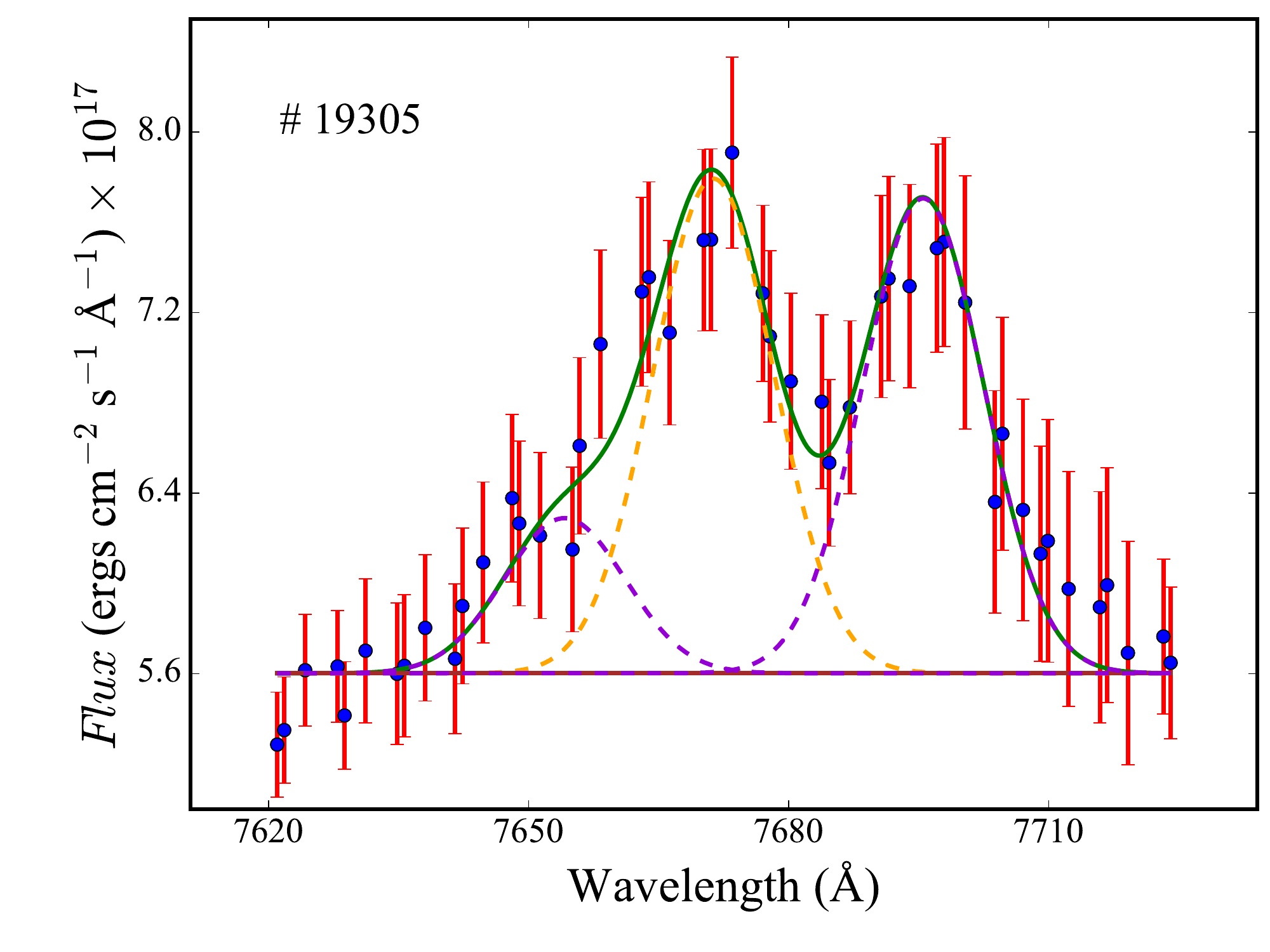}
 \includegraphics[width=0.36\textwidth]{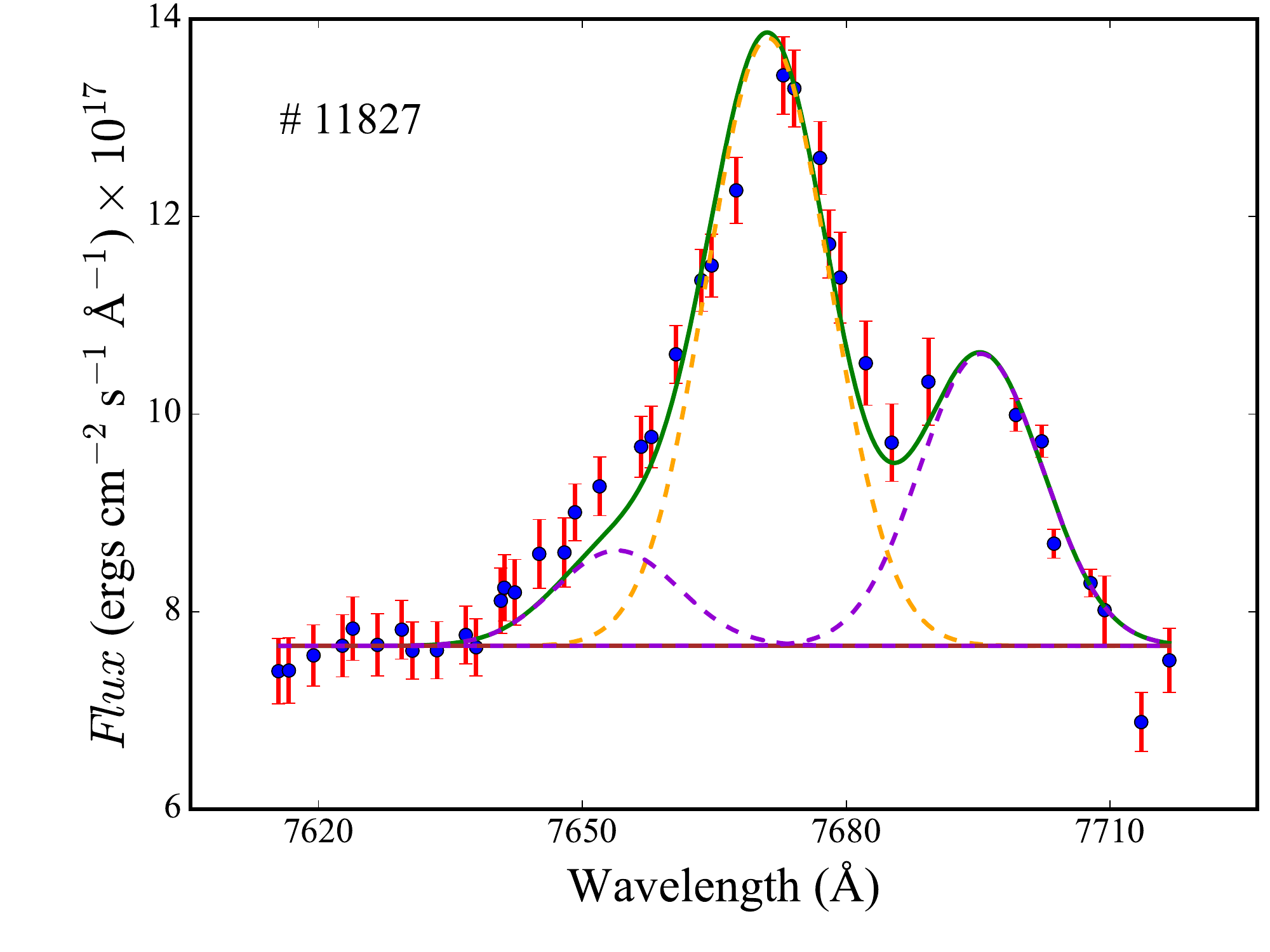}
\includegraphics[width=0.36\textwidth]{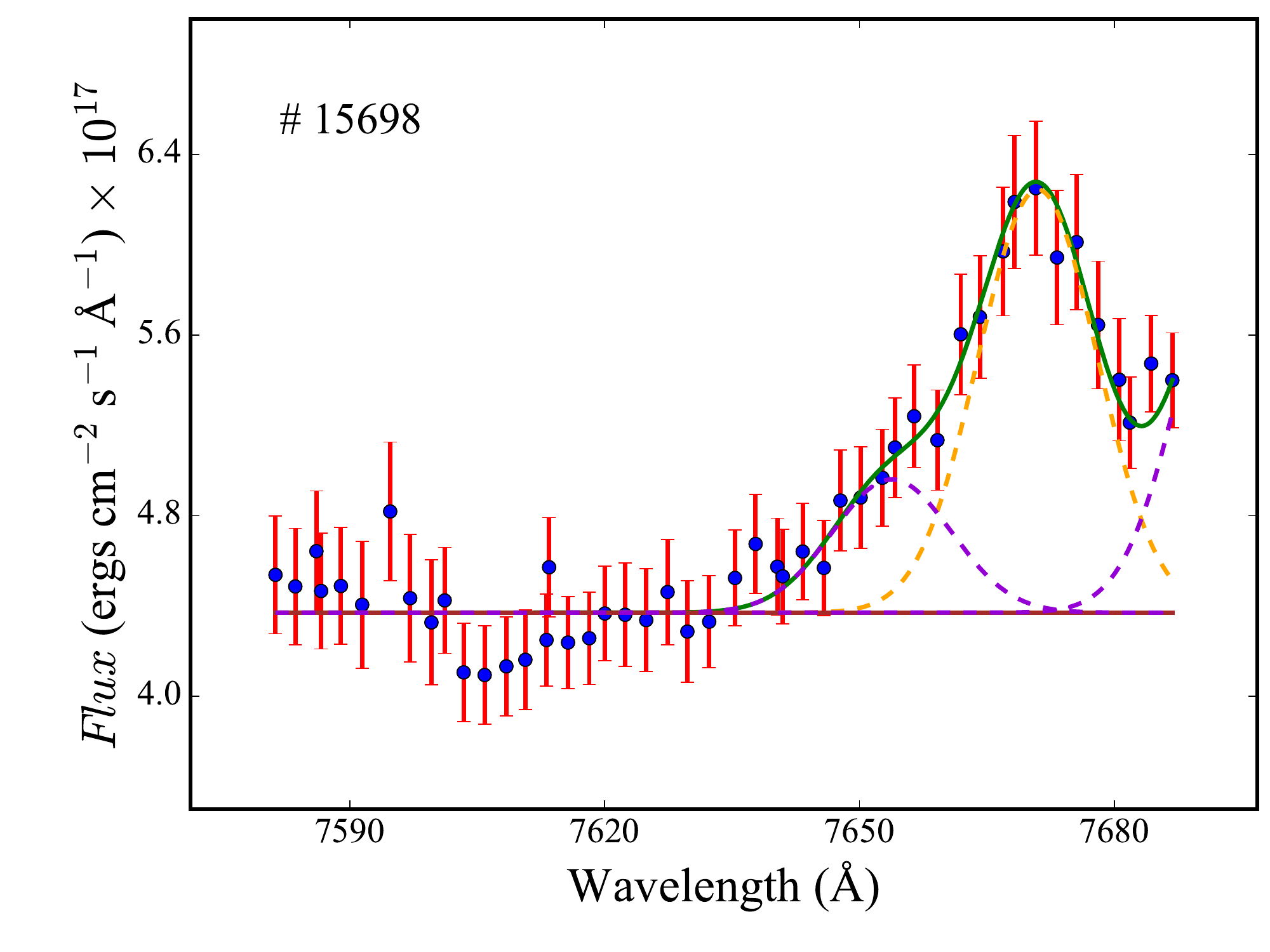}
\caption[AGN spectra]{OMEGA spectra of four of the X-ray-detected AGN in the A901/2 system (shown as crosses in Figure~\ref{WHAN}). 
In yellow and purple we show the fit to the \Ha and \NII lines, respectively, and in green the fit to the whole spectrum. Objects 9020 and 12953 are also identified as optical AGN with OMEGA data. Note the strong \NII lines. The galaxy 11827 has a very low formal probability of being an optical AGN, with its weaker \NII lines, although it is located just to the left of the boundary in the WHAN diagram (see Figure~\ref{WHAN}). Object 15698 doesn't meet the selection criteria because \NIIb is just outside the observed wavelength range. The \ratio ratio is therefore not well constrained for this object, although \NIIa appears reasonably strong (circled cross in Figure~\ref{WHAN}).  }
\label{fig:AGN_spectra}
\end{center}
\end{figure}

In their X-ray studies, \citet{Gilmour_2007} detected 12 galaxies as \textit{XMM} point sources in the A901/902 field (see their Table~12). These objects are present in our parent sample and we are able to produce OMEGA spectra for all of them. From these 12 objects, the two most luminous sources in the X-ray (12953 and 41435) have photometric redshifts (1.4 and 0.33, respectively) that prevent them from being included in our sample following the selection cuts Section (\ref{photometricredshiftcuts}), although the photometric redshifts of such strong AGN could be very uncertain.  

Of the remaining 10, only 3 have central OMEGA spectra which fulfil all our selection criteria (including the detection of \Ha and \NII at the system redshift). These three X-ray AGN have IDs 9020, 11827 and 19305, with formal probabilities of being an optical AGN of $100\%$, $1.1\%$ and $100\%$, respectively, according to the WHAN diagnostic diagram (see black crosses in Figure~\ref{WHAN}). Two of them (9020 and 19305) are therefore robust optical AGN. In the case of 11827, the \ratio ratio (which has very small formal errors given the quality of the data) places this galaxy very close to the vertical boundary between star-forming galaxies and AGN (but just to the left of the line; see Figure~\ref{WHAN}). Given the uncertainties in the exact location of this line (cf.~Section~\ref{sec:whandiag}), it is reasonable to conclude that it is quite possible that this object also shows signs of nuclear activity in the optical.  
The nuclear spectra of these three galaxies are shown in Figure~\ref{fig:AGN_spectra}. The prominent \NII lines are clear indicators of AGN activity in 9020 and 19305, while they are weaker in 11827. 

A fourth source (15698) is detected in \Ha and meets the redshift criteria, but it doesn't make the selection because \NIIb is just outside the observed wavelength range and therefore the \ratio ratio cannot be well constrained, although \NIIa appears reasonably strong. 
Its spectrum is also shown in Figure~\ref{fig:AGN_spectra}, and the location of this galaxy on the WHAN diagram (Figure~\ref{WHAN}) is indicated with a circled black cross. The available data suggest that this galaxy could also be an optical AGN, albeit with higher uncertainty.

In summary, at least two of the X-ray AGN (9020 and 19305) are classified as bona-fide optical AGN, with
the possibility of two more, 11827 and 15698, also being optical AGN.  This result implies that, out of the 12 X-ray AGN detected by \cite{Gilmour_2007}, at least two, and perhaps four galaxies (17--33\%) also show optical signatures of the presence of AGN. If we only consider the 10 X-ray AGN whose spectroscopic or photometric redshifts place them in the cluster structure, and thus within the detection limits of OMEGA, the fraction of X-ray AGN with optical AGN signatures could be as high as 20--40\%. \citet{Martini_2006} found that only 4 of at least 35 X-ray objects ($\sim11\%$) would be classified as optical AGN from their emission-line signatures, which is a smaller proportion than what we find. 

The lack of optical signatures in X-ray AGN might be a consequence of these objects being heavily obscured AGN and therefore they would not be expected to show optical signatures of nuclear activity. On the other hand, the fact that in the optical we detect about three times more AGN galaxies than in X-rays clearly shows that OMEGA is able to detect a large number of optical AGN with relatively weak X-ray emission. The four X-ray AGN we identify as secure or possible optical AGN are among the brightest \Ha emitters in the OMEGA-identified optical AGN sample (their \Ha luminosity is, on average, 2.5 times brighter than that of the OMEGA AGN population). The correlation that exists between X-ray luminosity and optical emission (albeit, with large scatter; see, e.g., \citealt{Lusso_2010,Stern_2012}), suggests that the vast majority of our optical AGN are probably too faint in X-rays to be detected in the available X-ray data.

\subsection{AGN properties}
\label{AGNprops}
We now explore the properties of the OMEGA AGN and SF galaxies and compare them to the overall cluster population. We carry out this analysis using two samples in parallel: one including only the robust detections (probability of being an AGN or a SF galaxy $\ge 99.7$\% or significance $\ge3\sigma$ from the WHAN diagram) and another one including also the objects that appear in either the \emph{Seyfert} or the \emph{star-forming} regions of the WHAN diagram, regardless of the significance of their classification. This increases the sample sizes by $\sim25$\% in a way that is statistically robust when analysing the properties of these two populations. In what follows we will see that the conclusions obtained using both samples are entirely compatible. 

\begin{figure}
\begin{center}
\includegraphics[width=0.48\textwidth]{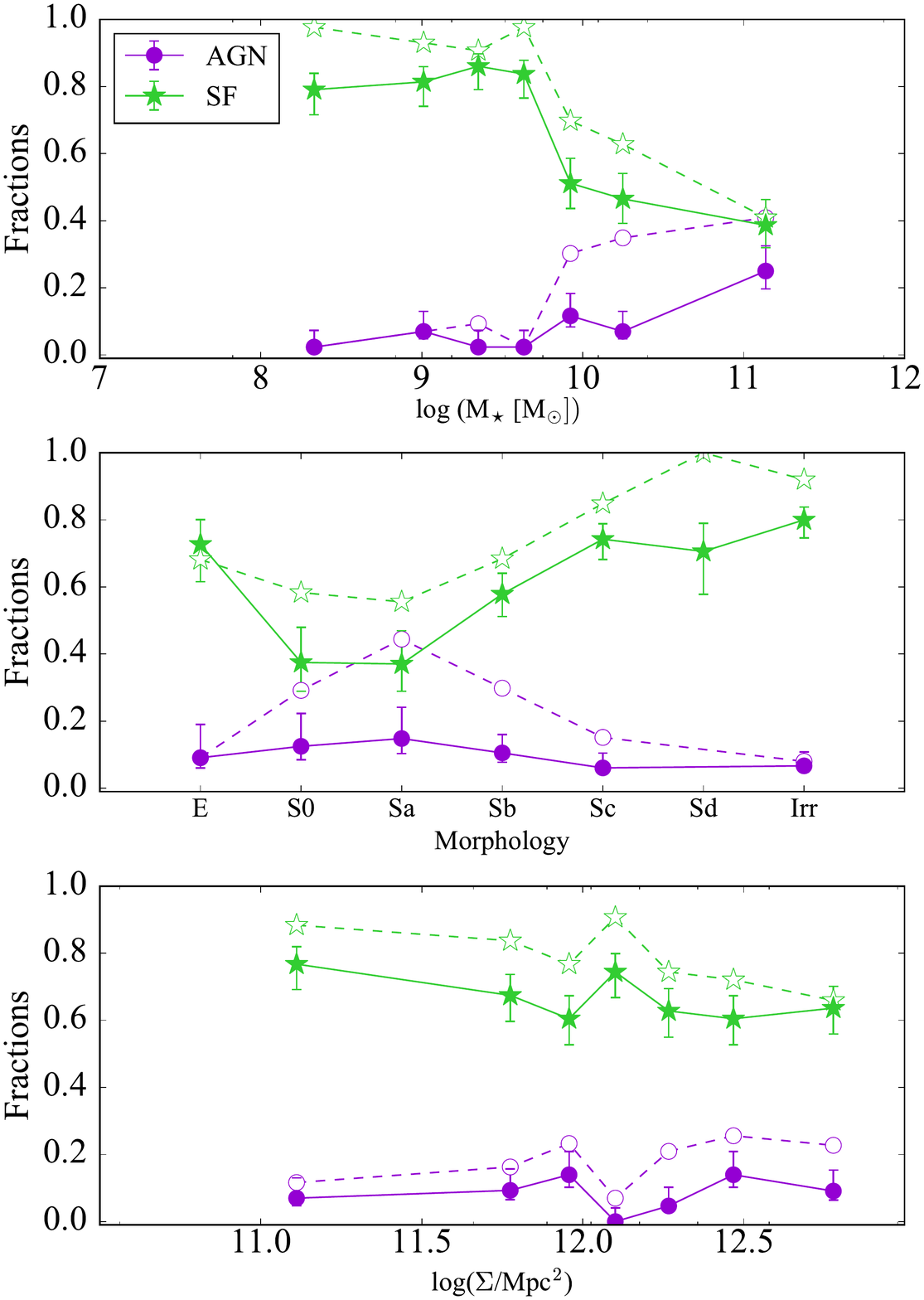}
\caption[]{Fractions of the cluster \Ha-emitting galaxies that are star-forming (green) or AGN (purple) as a function of the galaxies' stellar mass
(\emph{top}), morphology (\emph{middle}), and environmental stellar mass density (\emph{bottom}). Only objects where both \Ha and \NII have been observed in the central aperture ($R_{\rm PSF}$) spectra are used when computing these fractions. The filled symbols and solid lines correspond to galaxies with robust classifications ($\ge3\sigma$; see text for details). The open symbols and dashed lines represent the AGN and star-forming fractions including all objects in the corresponding region of the WHAN diagram. We only show error bars for the fractions of robust classifications for clarity. The stellar mass and environmental density bins are made to contain the same number of objects.}
\label{3histograms}
\end{center}
\end{figure}

The top two panels of Figure \ref{3histograms} present the star-forming and AGN galaxy fractions of the total cluster \Ha-emitting galaxy population as a function of stellar mass and morphology.  Only objects where both \Ha and \NII have been observed in the central aperture ($R_{\rm PSF}$) spectra are used when computing these fractions since both lines are needed to classify the galaxies. Optical AGN hosts span a wide range in masses although they are more frequent at higher masses. Moreover, these AGN appear in galaxies with all morphological types, with a small tendency to favour disk galaxies with prominent bulges (S0--Sb). 

The fraction of \Ha-detected elliptical galaxies that are classified as star-forming rather than AGN seems to be relatively high, but the uncertainty on this fraction is large since only $\sim10\%$ of all the elliptical galaxies are detected in \Ha (cf. Figure~\ref{morphofraction}). Taken at face value, it would imply that when \Ha emission is present in elliptical galaxies it is more often associated with residual star formation than with an optical AGN. However, it is also possible that because the \Ha emission detected in these galaxies tends to be quite weak, some of them may be misclassified LINERs.

\begin{figure}
\begin{center}
\includegraphics[width=0.48\textwidth]{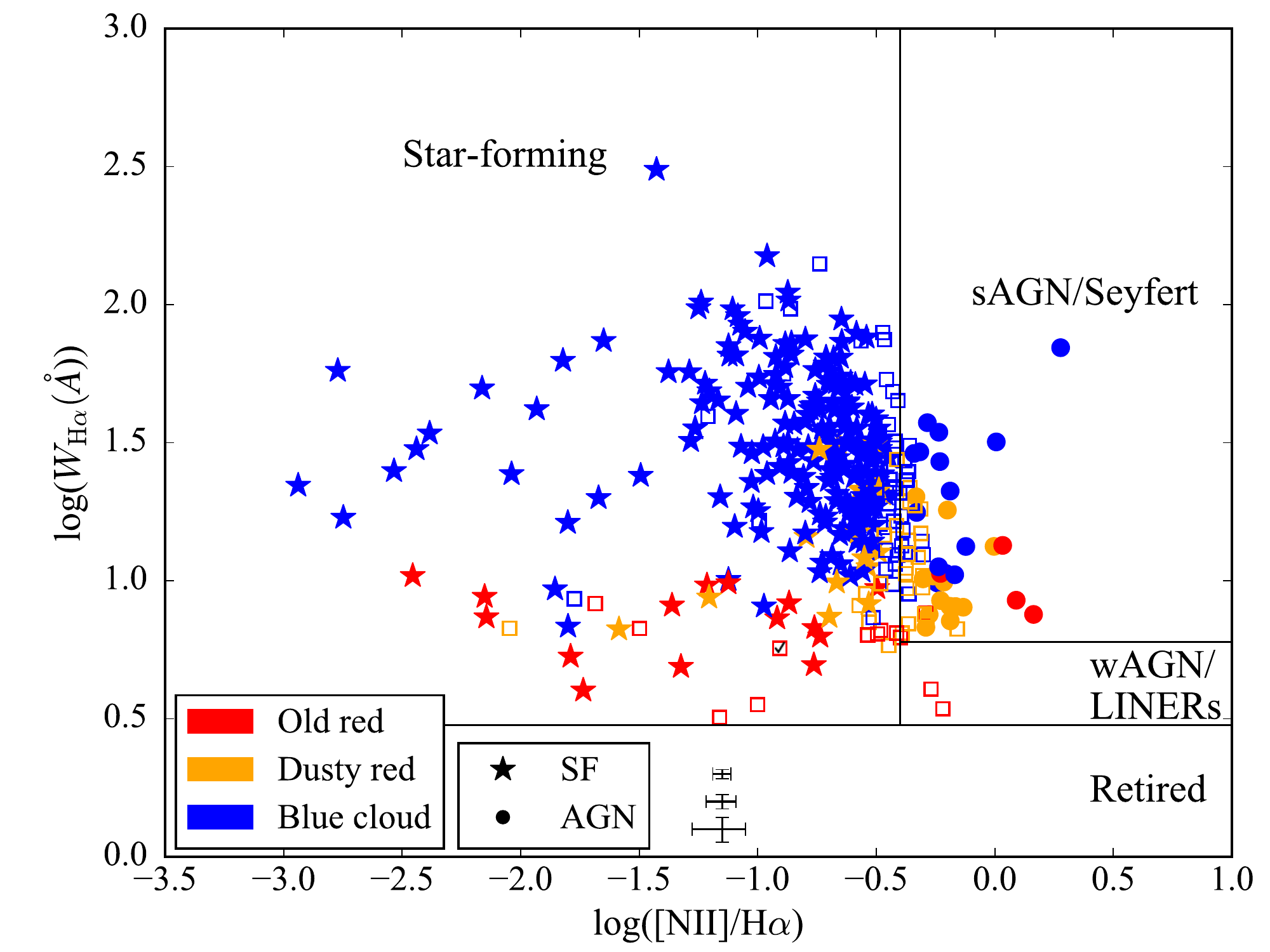}
\caption[]{WHAN diagram colour-coded by the SED type of the galaxies. In the bottom corner we show the 25, 50 and 75 percentiles of the errors for reference. Filled stars and circles correspond to galaxies robustly classified as SF and AGN, respectively, while empty squares represent galaxies that contain mixed sources of ionization and/or the classification is more uncertain.}
\label{seds_ewhan}
\end{center}
\end{figure}

Finally, we explore the SED types of the AGN population by plotting the WHAN diagram colour-coded by SED type in Figure \ref{seds_ewhan}. Whereas, as expected, most of the star-forming population belongs to the \emph{blue cloud} (86\%), there is a significant fraction of AGN that are classified as \emph{dusty reds} (42\%). This class of galaxies have reduced SFR and therefore tend to have lower \WHa. With the SF-related \Ha emission suppressed, any other source of ionization, including an AGN, would be easier to detect than in galaxies with more dominant star formation. This could explain their relatively high presence among the objects classified as AGN in the WHAN diagram. There are also a few galaxies classified as \emph{old red} among the low-\WHa population. This is not surprising because our deep \Ha imaging is able to detect very low levels of SF or AGN activity which are too weak to affect significantly the colours or SEDs of the galaxies.

\subsection{Environmental influence on AGN activity}
Apart from studying the properties of the objects hosting an AGN, our observations allow us to explore whether the presence of AGN is related to the local environment within the A901/2 system. The bottom panel of Figure \ref{3histograms} shows the fraction of \Ha-emitting galaxies that are AGN or star-forming as a function of environmental stellar mass surface density. Within the A901/2 cluster environment we find that the fraction of AGN remains roughly constant at all densities, with values always below $15\%$ of the \Ha-emitting population. 

The dependence of AGN activity on environment can be further evaluated by comparing the  \ratio line ratio distributions of the galaxy populations residing in the infall and core regions, as defined in Section~\ref{envdef}. In the WHAN diagram, this line ratio can be used to separate star-forming galaxies and AGN for \WHa$\,> 6$\AA. Therefore, rather than use a binary SF/AGN division we can use this line ratio as a continuous measure of AGN activity. This has the advantage of being independent on the exact location of the SF/AGN boundary in the diagram. 

\begin{figure}
\begin{center}
\includegraphics[width=0.45\textwidth]{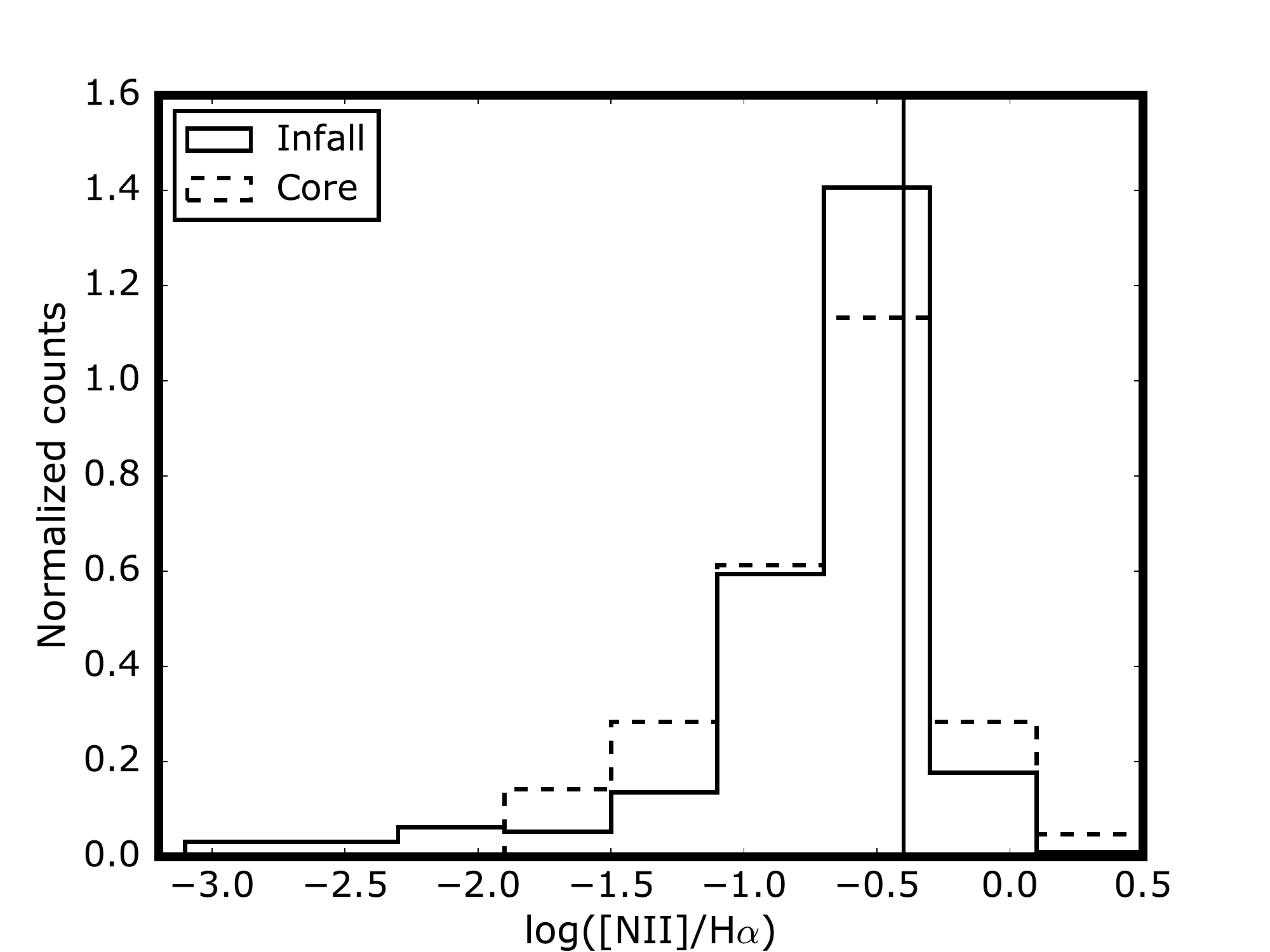}
\caption[\ratio line ratio distributions in the infall and core regions of the cluster.]{\ratio line ratio distributions for the \Ha-emitting galaxies with \WHa$\,>6$\AA\ in the infall (solid line) and core (dashed line) regions of the cluster system.  The vertical black line shows the fiducial separation between SF and AGN used in the WHAN diagram (Figure~\ref{WHAN}). }
\label{lineratio_agn}
\end{center}
\end{figure}

The distributions of \ratio in the infall and core regions for galaxies with \WHa$\,>6$\AA\ are shown in Figure \ref{lineratio_agn}. The \ratio distributions of the infall and core regions are quite similar. A K--S test gives a probability $\sim 67\%$ of them being drawn from the same population, which is consistent with the lack of environmental dependence of the AGN fraction we found before.

Our results are in contrast with the findings of \cite{Gilmour_2007}. They found evidence suggesting that the X-ray AGN in A901/2 tend to avoid the highest and lowest density regions. Given that the optical and X-ray AGN samples are very different, it is not surprising that they occupy regions of different environmental densities. This absence of environmental dependence agrees with what has been found in previous studies (\citealt*{Martini_2002}, \citealt*{Miller_2003}, \citealt*{Martini_2006}), although the lack of true field galaxies in our sample prevents a direct comparison with other works such as \citet{Kauffmann_2004} that found the fraction of high-luminosity optical AGN in the field to be around twice larger than in the cluster environment.

This diversity in the results found on the dependence of AGN activity on environment might be to some extent due to the use of different samples and indicators to probe AGN activity, together with inconsistent definitions of environment. In the case of A901/2 it would be extremely useful to have a comparable sample of \Ha-selected galaxies in the field in order to test whether there is any difference in the AGN content with respect to the cluster system. It would also be useful to have similar samples for a range of clusters with different masses and dynamical states.

\section{Integrated star-formation properties}
\label{sec:int_SF}
We now turn our attention to the study of the global star formation rates (SFRs) of the objects where we detect \Ha. To obtain the integrated SFRs we use the measurements within $R_{\rm tot}$, which encompasses the whole extent of the galaxies. 

\subsection{Star Formation Rates}
\label{subsection:sfrs}
In Paper~I we converted \Ha luminosities into SFRs using a constant conversion factor which included an averaged 1~mag correction for extinction. However, the effects of dust extinction change as a function of stellar mass \citep{Brinchmann_2004}, being of greater importance for more massive galaxies. Therefore, we use the results shown in Fig.~6 of \cite{Brinchmann_2004} to apply a mass-dependent dust attenuation correction to estimate SFRs from the \Ha luminosities. Thus, the original \citet{Kennicutt_1998} relation to convert \Ha luminosities into SFRs, takes the following form
\begin{equation}
\label{SFReq}
\mathrm{SFR}_{\mathrm{H}\alpha}\, = 7.9 \times 10^{-42} \times 10^{[A_{\mathrm{H}\alpha}(M_\star)/2.5]} L_{\mathrm{H}\alpha}\,\left[\mathrm{erg\,s}^{-1}\right]\;,
\end{equation}
in units of $\mathrm{M}_{\odot}\,\mathrm{yr}^{-1}$. In this equation, $A_{\mathrm{H}\alpha}(M_\star)$ is the stellar-mass-dependent extinction at the wavelength of \Ha. 

SFRs for the galaxies in A901/2 have also been calculated by \citet{Gallazzi_2009} using a combination of the extinction-corrected UV light  (specifically, the COMBO-17-derived rest-frame $2800$\AA{} luminosities) and the dust-reprocessed IR light (as observed with Spitzer/MIPS at $24\mu$m). In star-forming galaxies, the rest-frame UV light comes from massive stars and traces short star-formation timescales ($\sim10^{7}$--$10^{8}\,$years), albeit slightly longer than \Ha ($\sim10^{6}$--$10^{7}\,$years). UV light is much more strongly attenuated by dust than \Ha (by almost one order of magnitude). Dust absorbs the UV radiation coming from the stars, heats up, and re-radiates the energy in the far-IR, therefore tracing longer star formation timescales ($\sim10^{8}\,$years). 

The SFRs derived from the UV and IR are also available in the STAGES master catalogue \citep{Gray_2009}. Due to the different coverage of the COMBO-17 and Spitzer/MIPS observations, IR data are not available for all the objects. In these cases, only UV data was employed to estimate the SFRs, which are considered as lower limits since the amount of radiation re-emitted by the dust is unknown. When IR data are available but the measured flux is below the detection limit of $58\mu$Jy (corresponding to an IR-only SFR of 0.14 M$_{\odot}$yr$^{-1}$), the IR luminosities were predicted from UV-optical luminosities \citep[see Section 3.2 in][]{Wolf_2009}, representing very uncertain upper limits and thus not used here. When IR is detected, the SFRs were calculated using Equation 1 in \citet{Gallazzi_2009}. 

\begin{figure}
\begin{center}
\includegraphics[width=0.45\textwidth]{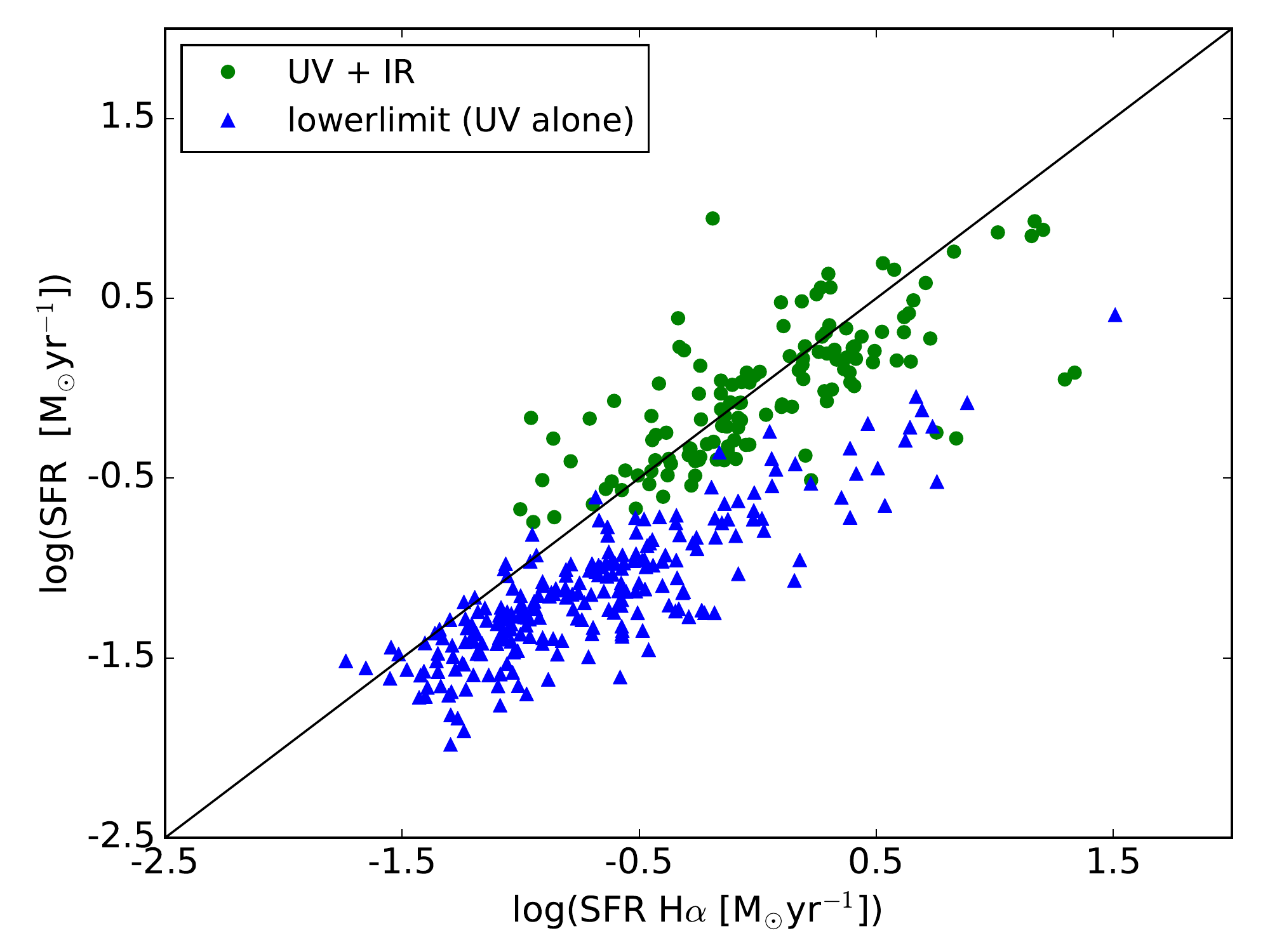}
\caption[Comparison of star-formation indicators]{Comparison of $\mathrm{SFR}_{\mathrm{H}\alpha}$, derived using OMEGA \Ha fluxes, with SFR estimates based on UV and IR photometry (see text for details). 
}
\label{SFRHa_UV_IR}
\end{center}
\end{figure}

In Figure \ref{SFRHa_UV_IR} we compare these multiwavelength SFRs with those estimated using OMEGA \Ha fluxes (Equation~\ref{SFReq}). In all cases, we use the cluster redshift ($z=0.167$) to calculate distances instead of individual photometric or spectroscopic redshifts, which are affected by peculiar motions and/or have larger uncertainties. In this figure, blue triangles show the SFRs as measured when only UV data were available (lower limits) and green circles are SFRs for galaxies with UV and IR detections. Note that because of the relatively shallow IR imaging available, low mass galaxies are absent in the UV$+$IR detected subsample. 
The blue points (UV-based lower limits) show a clear $\sim0.4$\,dex offset from the 1:1 relation. This is easily understood since the effect of extinction is significantly stronger in the UV than in the optical. 

The green points, which account for both obscured and unobscured star formation, show quite a good correlation with our \Ha SFRs and not much of an offset. This suggests that the average mass-dependent extinction correction applied to the \Ha luminosities works reasonably well, and, statistically, we recover the total SFR from the extinction-corrected \Ha. However, there is significant scatter ($\sim0.3$dex). Much of this scatter can be explained by galaxy-to-galaxy differences in dust content and/or extinction (recall that we apply extinction corrections that are averages for a given stellar mass). The different star-formation timescales measured by \Ha and the IR dust emission may also add some scatter. These issues will be explored in much more detail in subsequent papers.

\subsection{The \Ha Luminosity Function}

The galaxy luminosity function (LF) measures the number of galaxies with different luminosities in a given region of the Universe. The distribution of galaxy luminosities can provide relevant information about the evolutionary processes affecting galaxies. We present here the distribution of \Ha luminosities for the members of the A901/2 cluster structure, and compare it with other clusters and the field. 

\begin{figure}
\begin{center}
\includegraphics[width=.45\textwidth]{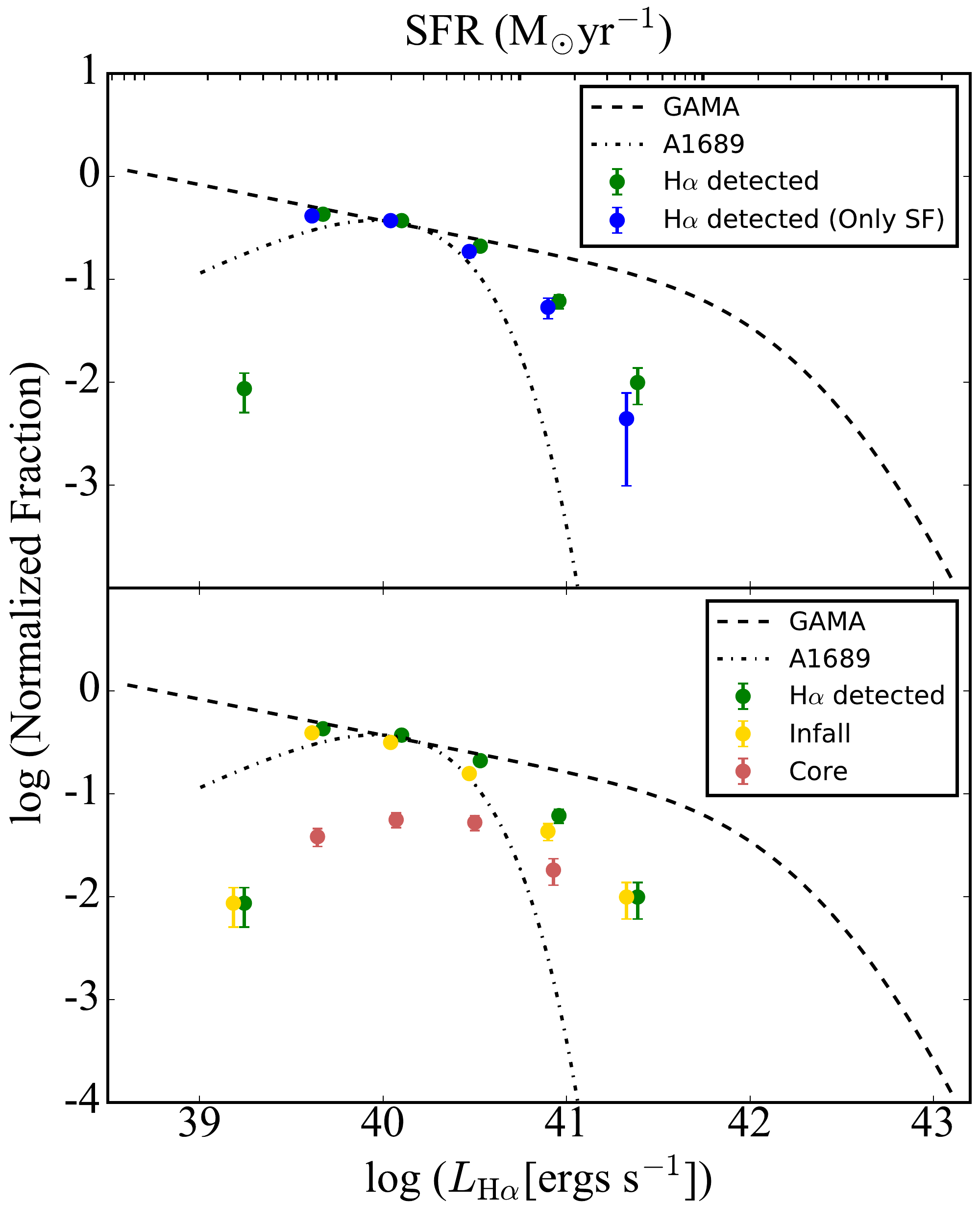}

\caption[Luminosity function]{The \Ha luminosity function for different subsamples of the OMEGA \Ha-detected galaxies. All the LFs are corrected for contamination and completeness. Errors are always Poissonian. \textit{Top panel:} LFs for galaxies in the whole A901/2 structure. The green points take into account all the \Ha-detected galaxies, while the blue points only consider the objects robustly ($>3\sigma$) classified as  star-forming galaxies using the WHAN diagram. The green and blue points have been slightly shifted horizontally for clarity. \textit{Bottom panel:} Separate LFs for \Ha-detected galaxies in the whole A901/2 structure (green points), the core region (brown points) and the infall region (yellow points). The green and yellow points have been slightly shifted horizontally for clarity. In both panels we show for comparison the \Ha LF of the A1689 cluster at $z=0.18$ from \citet[][black dash-dot line]{Balogh_2002} and the LF of the field at $z=0.1$--$0.2$ by \citet[][black dashed line]{Gunawardhana_2013} from the GAMA survey. 
The scale on the top axis shows SFRs calculated from \Ha luminosities using Equation~\ref{SFReq} with an extinction correction $A_{\mathrm{H}\alpha}\simeq1\,$mag (corresponding to a galaxy with stellar mass $3\times10^9$M$_{\odot}$ the median stellar mass of the sample).  }
\label{lumfunc}
\end{center}
\end{figure}

In Figure~\ref{lumfunc} (top panel) we show the \Ha LF for all \Ha-detected galaxies (dark green points) and for those galaxies robustly classified as star-forming objects (blue points; cf. Section~\ref{sec:whandiag}). The top axis shows SFRs estimated with Equation \ref{SFReq} and applying an extinction correction of $A_{\mathrm{H}\alpha}\simeq1\,$mag, which corresponds to a galaxy with mass $3\times10^9$M$\odot$ (the median stellar mass of the sample). These LFs have been corrected for incompleteness using the empirical detection fractions determined in Section~\ref{completeness} for each galaxy type and $R$-band magnitude (cf. Figure~\ref{Rmag}; a linear interpolation is used across $m_R$ bins). The OMEGA \Ha LF reaches well below $0.1$M$_\odot\,$yr$^{-1}$. Furthermore, the \Ha LF for all \Ha-detected galaxies is very similar to the one derived for the star-forming galaxies only at almost all luminosities: many of the green and blue points are almost indistinguishable. This indicates that SF galaxies dominate the \Ha emission of the galaxy population at almost all luminosities. However, at the highest \Ha luminosities the contribution from AGN to the LF becomes noticeable (AGN live preferentially in brighter, more massive galaxies; see Section~\ref{AGNprops}). For the lowest luminosity point, the SF/AGN classification becomes too uncertain, and the incompleteness correction is probably too unreliable for this point to be taken seriously.

For comparison with \Ha LFs in other environments at similar redshifts, we show the LF of the A1689 cluster ($z=0.18$) from \citet{Balogh_2002} and that of the field from the GAMA survey \citep{Gunawardhana_2013}. The field \Ha LF shown corresponds to the best-fitting parameters of the Saunders function (\citealt{Saunders_1990}) that \citet{Gunawardhana_2013} obtain for the $z = 0.1$--$0.2$ redshift bin  ($\log{L^{*}} = 34.55$, $\log{C} = -2.67$, $\alpha = -1.35$, $\sigma = 0.47$). The A1689 LF uses the \citet{Schechter_1976} function parameters given by \citet[][i.e., $\alpha=-0.1$ and $L^{*}=10^{40}\;\mathrm{erg\,s}^{-1}$]{Balogh_2002}. These two \Ha LFs use the same cosmology as we do. All the \Ha LFs are normalized such that they have the same value at $L_{{\rm H}\alpha} = 1\times10^{40}$ergs~s$^{-1}$.  
The main conclusion of this comparison is that the OMEGA \Ha LF for the A901/2 cluster structure is intermediate between that of the field, where environment has not yet played any star-formation-quenching role, and the more evolved and dense cluster A1689, where galaxies have been processed by the environment for a longer period of time than in A901/2. 

Although the statistical uncertainties become quite large when analysing the LF of core galaxies separately from that of the infall region, the bottom panel of Figure~\ref{lumfunc} shows that the LF in the core of A901/2 (brown points) seems more similar to that of A1689 than that of the A901/2 structure as a whole. In particular, there are no galaxies in the core with \Ha luminosities above $10^{41}\,$erg s$^{-1}$.

\subsection{Effects of mass and environment on the galaxies' star-formation properties}
\label{massenveffect}
 
The variation of star formation activity as a function of environment has been widely discussed in previous works. Some claim that the average SFR in star-forming galaxies remains roughly constant with environment, implying that it is only the fraction of SF galaxies what changes \citep{Balogh_2004_ecology,Verdugo_2008,Bamford_2008}. On the contrary, other studies have detected declining SFRs towards denser regions: \citet{Poggianti_2008,Vulcani_2010,Paccagnella_2016}. The timescale of the transition  from being actively star-forming galaxies to becoming passive in these two scenarios is completely different. A change in the fraction of star-forming galaxies without detecting a change in their SFRs implies a rapid transition from active to passive; however, this transition is much slower if there is a decline in the SFRs of star-forming galaxies.

\begin{figure*}
\begin{center}
\includegraphics[width=0.9\textwidth]{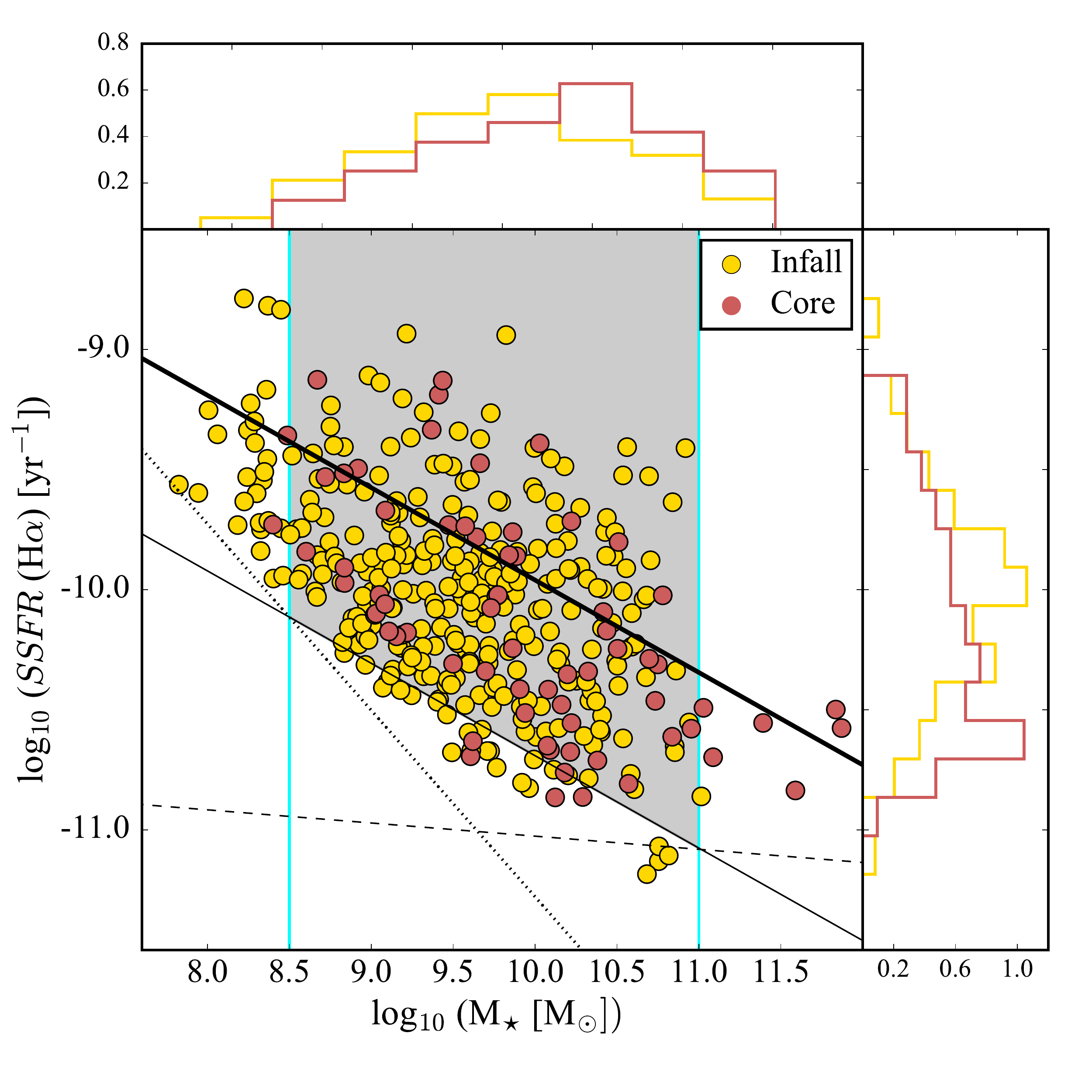}
\caption[SSFR vs mass]{SSFR vs stellar mass for the \Ha-detected galaxies in the infall (yellow) and core (brown) regions of the A901/2 system. The black solid line corresponds to the SSFR--stellar mass relation for the SDSS field galaxies. The dotted and dashed lines show the OMEGA survey \Ha flux and equivalent-with detection limits. The thin black solid line marks the lower boundary of the region free from incompleteness. This lower boundary, together with the cyan vertical lines, define the grey-shaded region containing an unbiased mass-limited sample. See text for details. We also show the stellar mass and SSFRs distributions of the core and infall samples as histograms. }
\label{allssfrmass}
\end{center}
\end{figure*}

To explore how the environment affects SF activity in A901/2 galaxies, we plot in Figure~\ref{allssfrmass} the specific star-formation rates (SSFRs; SFR per unit stellar mass) of the \Ha-detected galaxies as a function of their stellar mass. Objects that reside in the infall and core regions, as defined in Section~\ref{envdef}, are shown as yellow and brown points respectively. 
In order to understand this diagram, it is very important to take into account the detection limits of the OMEGA survey. As discussed in Section~\ref{completeness}, our ability to detect \Ha emission depends simultaneously on the \Ha flux and equivalent width. The dotted line in Figure~\ref{allssfrmass} corresponds to an \Ha flux detection limit of $F_{{\rm H}\alpha} = 3 \times 10^{-17}$erg$^{-1}$cm$^{-2}$s$^{-1}$and an average $\simeq1\,$mag extinction.  The dashed line shows the approximate \WHa detection limit corresponding to \WHa = 3\AA. In order to estimate the location of this line, stellar masses need to be converted into $R$-band luminosities (and thus continuum fluxes) using the mass-to-light ratios of the galaxies. The mass-to-light ratio varies with galaxy mass and SED type, but for the sake of simplicity we only show the limit corresponding to the \emph{dusty red} galaxies. If we used the equivalent line for the \emph{blue cloud} galaxies, which is marginally higher, our conclusions would not be affected significantly\footnote{Note that this line is not completely horizontal (the mass-to-light ratio depends weakly on galaxy stellar mass or luminosity), but its slope is quite shallow. This implies that an \WHa-selected sample is not far from a SSFR-selected one.}. We have detected very few galaxies below the dashed and dotted lines, suggesting that our estimates of the detection limits are reasonable.

Given that our sample only contains galaxies in the A901/2 cluster structure, to explore whether environment has altered the galaxies' SSFRs, it is useful to compare them with a field sample. For that purpose we use the SSFR--stellar mass relation derived from the SDSS Data Release~7 \citep[DR7;][]{Abazajian_2009}. These SFRs have been calculated in a similar way to ours (see section \ref{subsection:sfrs}), and were corrected to total SFRs by taking into account the fraction of the light outside the fibres and the colour gradients \citep{Brinchmann_2004, Salim_2007}. They are therefore suitable for direct comparison with our data\footnote{As a sanity check, and to test whether aperture effects in the determination of the galaxies' SFRs would affect our conclusions, we have repeated the analysis described below using SSFRs determined inside the SDSS fibres and OMEGA PSF \Ha fluxes. Our conclusions do not change.}. This relation is shown as a thick black solid line in Figure \ref{allssfrmass}. The median redshift of the SDSS sample used here ($0.106$) is sufficiently close to that of the A901/2 system that redshift evolution of the SFR main sequence should have minimal effect on our conclusions: following \cite{Whitaker_2012}, the main sequence would move up by only $\sim0.06\,$dex from $z=0.106$ to $z=0.167$. Such movement would, if anything, make some of our conclusions marginally stronger.

Without any further analysis, we find that most of the OMEGA \Ha-detected galaxies lie below the field relation at all stellar masses. 
This result indicates that a large fraction of the cluster galaxies, including those in the infall regions, have significantly reduced their SSFRs with respect to the field. Our findings therefore strongly support the scenario where galaxies falling into regions of higher density experience a relatively slow decline in their star-formation activity, i.e., a slow transition from star-forming to passive.
However, because \Ha is sensitive to star formation on a very short timescale, it would be necessary to look in detail at other SFR estimators such as UV or IR light, sensitive to longer timescales, to obtain a more complete picture. We will do this in a future paper (Wolf et al.\, in prep.)

Figure~\ref{allssfrmass} illustrates clearly why some studies may find that there is no environmental dependence on the average SSFR of star-forming galaxies, erroneously claiming that there is only a change in the relative fraction of star-forming and non-star-forming galaxies in different environments. The depth of the survey and its ability to find galaxies with low SFR/SSFR plays a crucial role here. A relatively shallow survey may be unable to identify as star-forming those galaxies with relatively low SSFR, wrongly implying that only the fraction of star-forming galaxies changes, but not their average SSFR. Because we are able to detect very low \Ha fluxes and equivalent widths (and thus very low levels of star formation), our sample contains \Ha-emitting galaxies that would have been missed by many surveys. Therefore, strongly suppressed galaxies still appear as \Ha emitters in our sample. {The presence of these strongly suppressed but still star-forming galaxies in our sample provides strong evidence that is not only the fraction of star-forming galaxies what changes with environment, but also the average SSFRs of the galaxies.}

The SSFR vs.\ stellar mass plot (Figure~\ref{allssfrmass}) provides a qualitative picture of star formation suppression in the A901/2 system. Our data also allow us to obtain quantitative information on how galaxies are evolving within the cluster environment, in particular, how much their SF activity is affected. To carry out such analysis, we need to work with a sample that is not biased by incompleteness due to $F_{{\rm H}\alpha}$ or \WHa selection effects. The selection of this sample is shown as a shaded region in Figure \ref{allssfrmass}. This region is defined in such a way that the left and right boundaries are mass limits (vertical cyan lines at $\log_{10}(M^*/{\rm M}_{\odot})=8.5$ and $11.0$) and the lower boundary (solid thick black line) is a diagonal line parallel to the SDSS field relation (solid thick black line) which is above the \Ha flux and equivalent width selection limits (dotted and dashed lines) for the selected stellar mass range. There is no upper bound to the selection box since galaxies with SSFR above the field line are not affected by the selections limits, and should therefore be included in the sample.  We require the lower boundary to be parallel to the SDSS field relation so that, at all masses, we are able to measure similar amounts of SF suppression (or vertical distance from the field line). We also require the sample to be stellar-mass limited. The stellar-mass limits are chosen arbitrarily, but in such a way that the selected galaxy sample is as large as possible while covering a wide range of stellar masses.

\begin{figure}
\begin{center}
\includegraphics[width=0.44\textwidth]{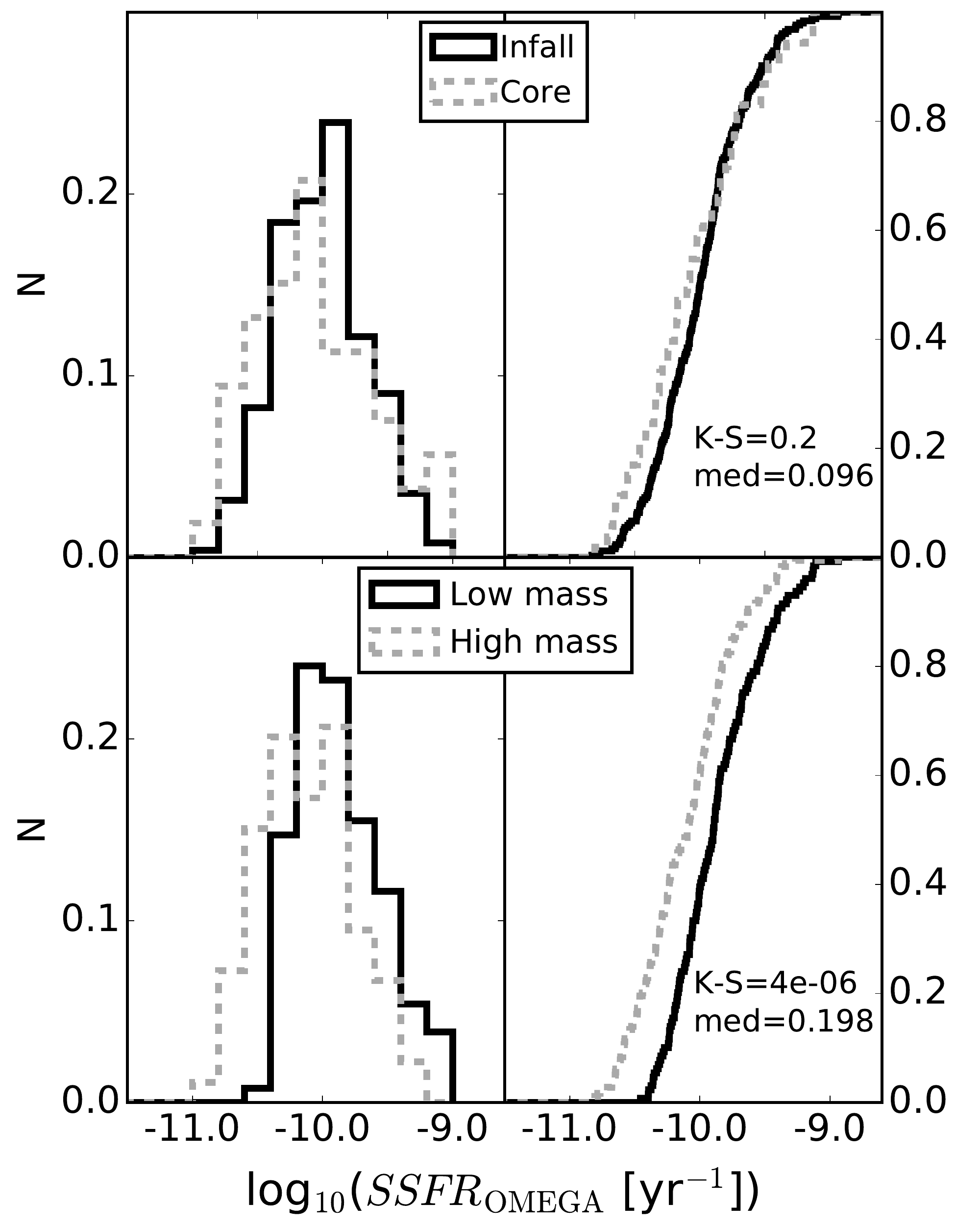}
\caption[]{Differential and cumulative histograms showing the SSFR distributions of galaxies split by environment (infall and core regions; top) and mass (split at $3\times 10^{9}$M$_{\odot}$; bottom). }
\label{SSFRmassdensity}
\end{center}
\end{figure}

Making use of this newly defined unbiased sample, we explore in detail the effects of mass and environment on the integrated star formation of the galaxies. In Figure \ref{SSFRmassdensity} we compare the SSFRs distributions of galaxies split by environment (infall and core regions) and mass (split at $3\times 10^{9}$M$_{\odot}$, the median of the stellar mass distribution). The median SSFR of the galaxies in the infall and core regions only differs by $\sim0.096\,$dex, and the distributions are very similar: a Kolmogorov--Smirnov (K--S) test yields a  probability $>23\%$ for both samples being drawn from the same population. For low- and high-mass galaxies, the change in the median SSFR is $\sim0.197\,$dex, with the K--S test yielding much lower probabilities ($\sim4\times10^{-6}$) for both distributions being the same. These results indicate that although both environment and stellar mass could drive variations in the SSFRs of the galaxies, the latter one seems to dominate within the limitations of the OMEGA sample. 

\begin{figure}
\begin{center}
\includegraphics[width=0.44\textwidth]{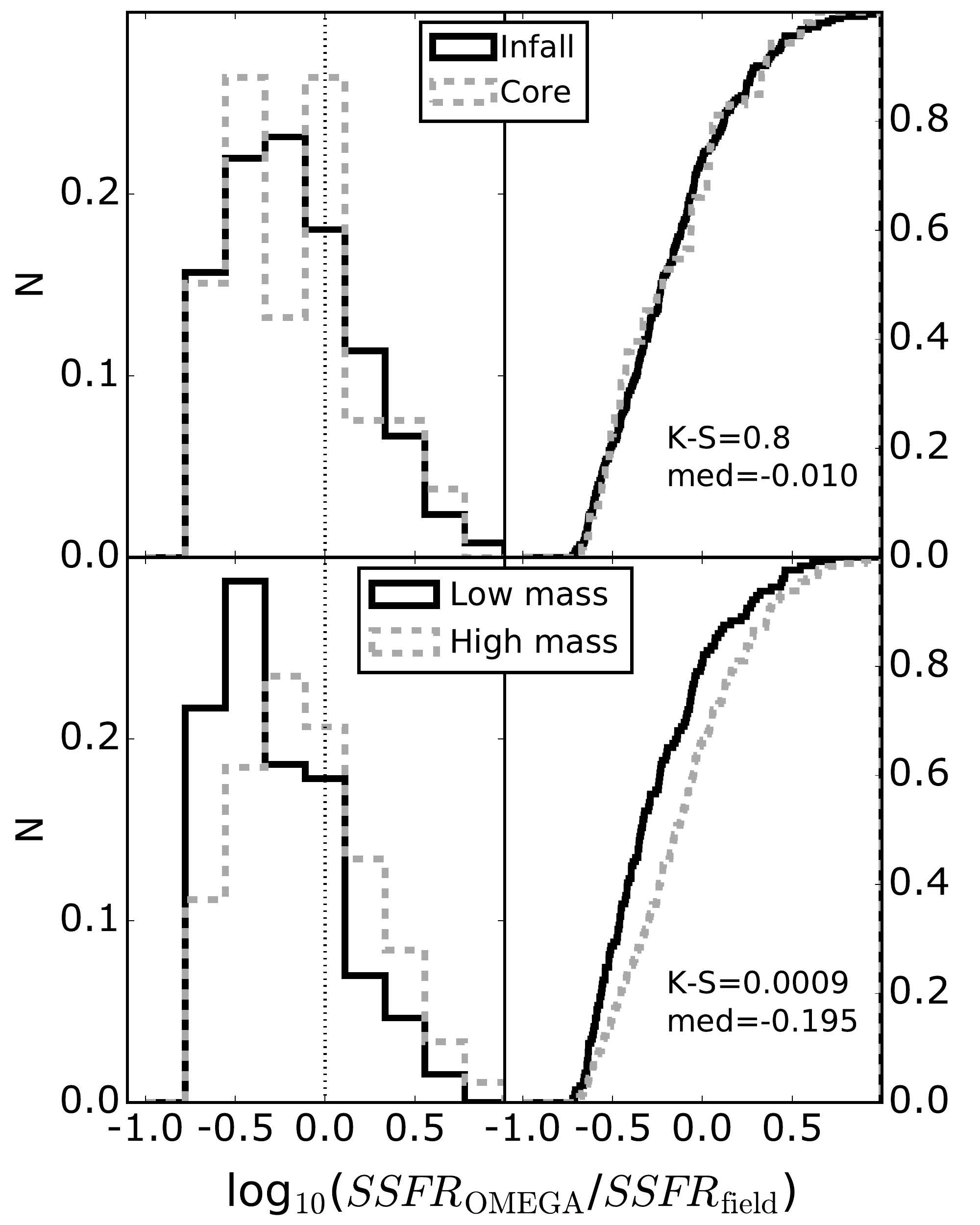}
\caption[Offset to the field relation]{The same as Figure \ref{SSFRmassdensity} but now using $\log_{10}$(SSFR$_{\rm OMEGA}$/SSFR$_{\rm field})$, which measures the SSFR suppression (see text for details). The vertical dotted line corresponds to the field relation, therefore objects to the left of this line are suppressed.}
\label{deficitmassdensity}
\end{center}
\end{figure}

However, this test is not entirely fair because stellar mass and SSFR are correlated, and the sample selection boundary is not horizontal. Taking into account that low-mass galaxies tend to have higher SSFRs than high-mass ones, to quantify the effect of the environment on the galaxies' SSFR it is better to use a measure of the \emph{SSFR suppression}. The vertical logarithmic distance from the field SSFR--mass relation would directly measure by what factor the SSFR of a galaxy of a given stellar mass has been suppressed with respect to the value it ``should have'' if it were in the field \citep{Whitaker_2012, Speagle_2014}. We will therefore use the quantity $\log_{10}$(SSFR$_{\rm OMEGA}$/SSFR$_{\rm field})$ to measure SSFR suppression. In Figure \ref{deficitmassdensity} we repeat the previous analysis but now using this quantity instead of SSFR.  As before, we find no difference in the SSFR suppression distributions of the core and infall galaxy samples. But there is a clear difference between high- and low-mass galaxies: the SSFR is suppressed more efficiently in low-mass galaxies than in high-mass ones. We therefore conclude that the SSFR of galaxies within the A901/2 structure tends to be suppressed regardless of whether they live in the core or the infall regions, but the suppression is significantly larger for the low-mass half of the sample (by $\sim60$\%). Thus, for the limited range of environments sampled by the OMEGA survey, the effect of stellar mass on star-formation activity is stronger than that of the environment. The A901/2 structure environment is able to suppress star formation more effectively on low-mass galaxies than in high-mass ones.

\begin{figure*}
\begin{center}
\includegraphics[width=0.9\textwidth]{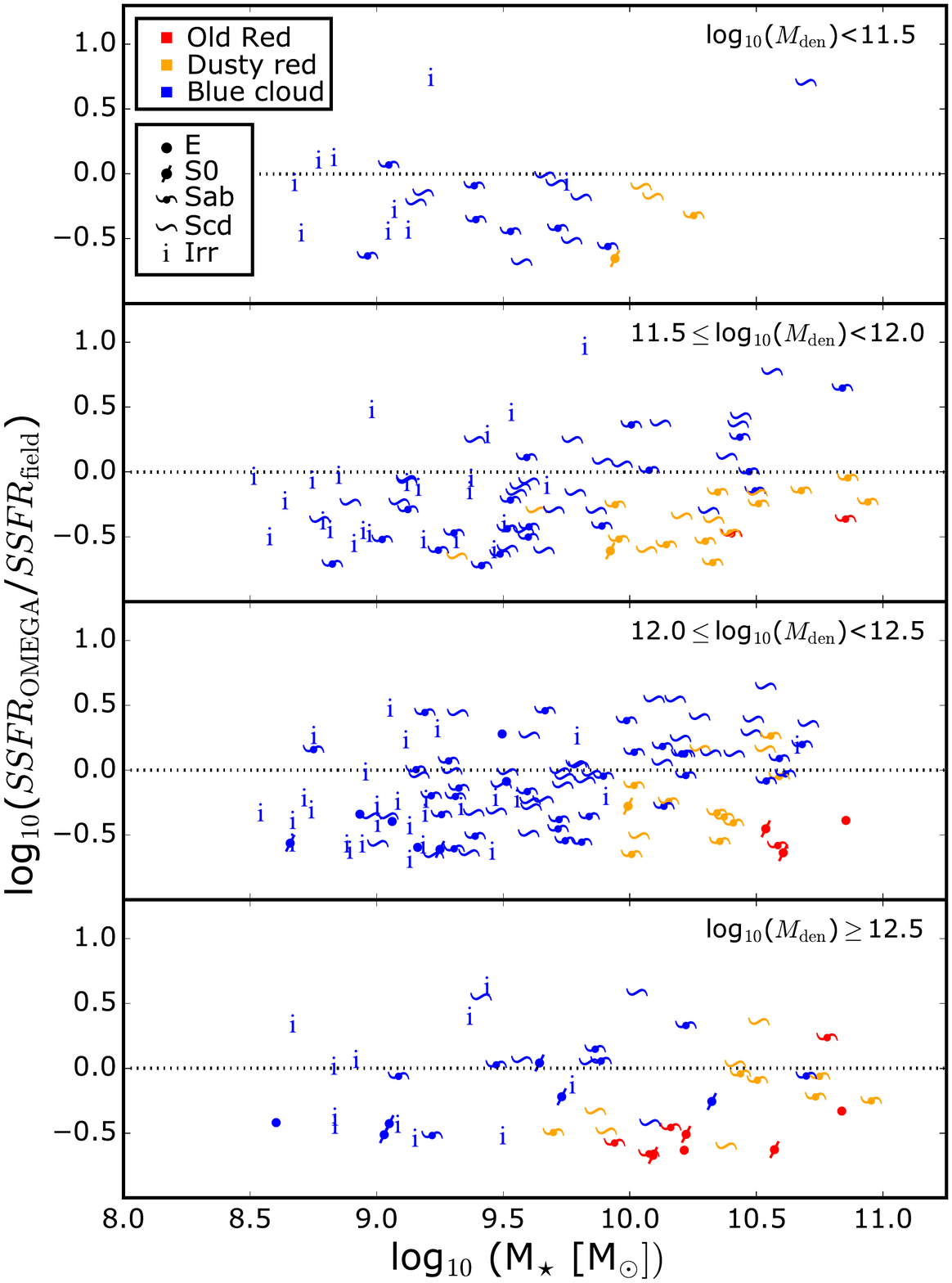}
\caption[SSFR vs mass different SED and morphological types]{SSFR suppression measure, $\log_{10}$(SSFR$_{\rm OMEGA}$/SSFR$_{\rm field})$, for the unbiased galaxy sample as a function of stellar mass. The sample has been split into four environmental stellar mass density bins. Symbols and colours correspond to different morphologies and SED types, respectively, as explained in the legend. The horizontal dotted line corresponds to the field relation. The further below this line a galaxy appears, the stronger its SSFR has been suppressed with respect to a field galaxy of the same stellar mass. 
}
\label{deficit4on1}
\end{center}
\end{figure*}

\subsection{Star-formation properties of galaxies with different morphologies and SED types}
\label{morphSEDeffect}

We move now to exploring how the galaxies' SSFR, SED type and morphology are linked with environment within the A901/2 system. In Figure \ref{deficit4on1} we plot our SSFR-suppression measure, $\log_{10}$(SSFR$_{\rm OMEGA}$/SSFR$_{\rm field})$, as a function of stellar mass, splitting the sample into four different density bins.  
Symbols and colours correspond to different morphologies and SED types, as explained in the legend. The dashed line corresponds to the field relation. The further below this line a galaxy is, the stronger its SSFR has been suppressed with respect to a field galaxy of the same stellar mass. Note that we use here the unbiased galaxy sample defined in Section~\ref{massenveffect}.

The data shown in Figure~\ref{deficit4on1} indicates that suppression of star formation with respect to the field happens at all the environmental stellar mass densities probed by our study with similar efficiency. We therefore confirm that, for the range of environments probed by the OMEGA sample, the quenching efficiency of the environment is similar. These results suggest that a significant fraction of the galaxies in the A901/2 structure may have experienced some environmental effects or ``preprocessing" before falling into the structure we observe now \citep{Zabludoff_Mulchaey_1998,McGee_2009, Haines_2015}, while the additional effect of the A901/2 environment is either negligible or roughly independent of the local environment within the system.

\begin{figure}
\begin{center}
\includegraphics[width=0.49\textwidth]{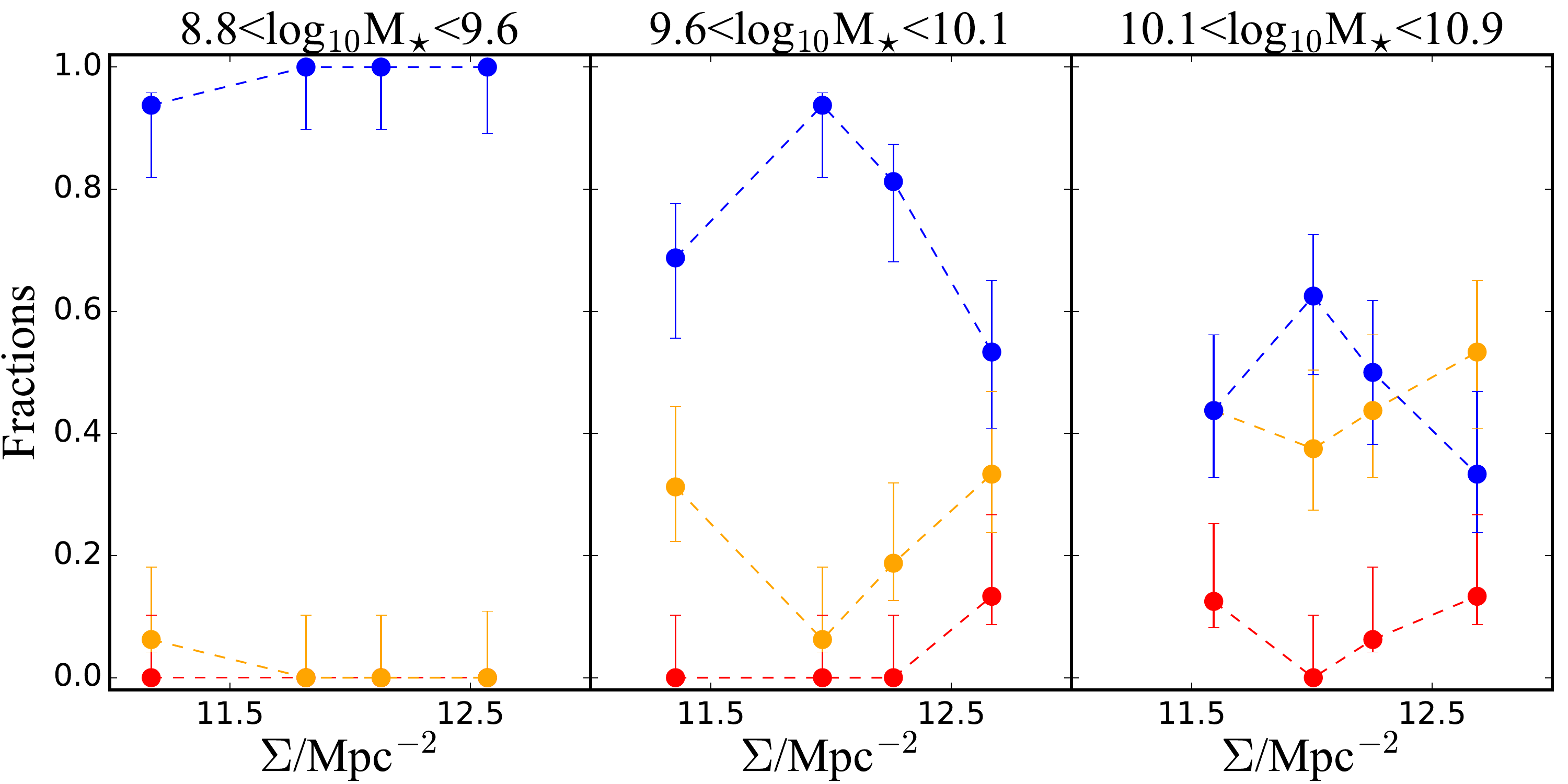}
\caption[Fraction of spiral galaxies with different SED types]{Fractions of \Ha-emitting spiral galaxies with different SED types as a function of environmental stellar mass density for three different stellar mass bins: \emph{blue cloud} (blue), \emph{dusty reds} (orange) and \emph{old reds} (red). The  \emph{dusty reds} phase become as common as the \emph{blue cloud} among star-forming galaxies with masses between $\sim10^{10}$ and $10^{11}\,$M$_{\odot}$. Bins are chosen to contain the same number of galaxies.}
\label{spiralseds}
\end{center}
\end{figure}

Most of the \Ha-detected galaxies shown in Figure \ref{deficit4on1} are spirals, although there is a relatively high number of galaxies with irregular morphologies at low masses. Interestingly, there is a high fraction of \Ha-emitting spiral galaxies with stellar masses $\geq1\times10^{10}\,$M$_\odot$ that are classified as \emph{dusty red} based on their SED type\footnote{Note that these galaxies are star-forming red spirals. See Section \ref{sec:stagesdata} for a discussion on the nature of the so-called \emph{dusty red} galaxies.}. Previous works by \citet{Bamford_2009} and \citet[][also in A901/2]{Wolf_2009} have shown that the fraction of spiral galaxies with \emph{dusty red} SEDs increases towards higher densities. To explore whether the same happens in our sample of \Ha-detected galaxies, we show in Figure~\ref{spiralseds} the fraction of \Ha-detected spiral galaxies with different SED types for different galaxy mass bins as a function of environmental stellar mass density. We find that star-forming \emph{dusty red} spirals become as common as \emph{blue cloud} ones for masses between $\sim10^{10}$ and $10^{11}\,$M$_{\odot}$. Therefore, in A901/2 we detect suppression of star formation at all masses, but at high masses the spiral galaxies enter a \emph{red} phase characterized by suppressed but still relatively high SFRs (hence implying that \emph{red} does not necessarily mean \emph{non-star-forming}) on their way towards denser regions. These findings agree nicely with the evidence for a slow transition from SF to non-SF found in Section \ref{massenveffect}, since if the transition was fast, a suspected intermediate type (\emph{dusty red}) should not occur in such high numbers.

Environmental effects such as ram-pressure stripping \citep{Gunn_Gott_1972,Yara_2015} or starvation \citep{Peng_2015} are therefore potential candidate processes to be responsible for the suppression of star formation in these disk galaxies since their SSFR and gas are affected \citep{Boesch_2013, Boesch_2013b} while retaining their overall morphologies.

\section{Summary and Conclusions}
\label{sec:summary}

In this paper we present a study of the integrated properties of the \Ha-emitting galaxies from the OMEGA survey. Based on Fabry--P\'erot observations obtained with the OSIRIS instrument at the $10.4\,$m GTC telescope, we have analysed the AGN activity and integrated star-formation properties of the galaxies in the multi-cluster system A901/2 at $z\sim0.167$. This analysis covers the whole survey area, and is therefore an extension of the work presented in Paper~I, which focused on the two densest regions covering only $\sim1/10$ of the observations. 

The survey design, observations, and preliminary data reduction are described in Paper~I. Building on the data reduction and analysis techniques presented there, we have further improved the sky subtraction and wavelength calibration based on the methods proposed by \citet{Weinzirl_2015}. After careful selection, we have built a sample of 439 galaxies with \Ha detected in their integrated spectra, of which 321 also have \NII emission lines in the observed wavelength range. This sample is used to study the integated star-formation properties of the galaxies. We have also built a second sample of 360 galaxies with \Ha and \NII emission detected in nuclear $R_{\rm PSF}$-aperture spectra, which we use to study AGN activity. All these galaxies belong to the A901/2 cluster system, with velocities covering $\simeq\pm2000\,$km$\,$s$^{-1}$ around the system's redshift. 
 
Using AGN diagnostics based on the \Ha and \NII emission lines, we find 33 robust AGN candidates. Cross-correlating our optical AGN sample with an X-ray-detected AGN sample covering the same region and containing 12 sources, we found at least two (and, likely, four) objects which are both optical and X-ray AGN. We attribute the lack of optical AGN counterparts of the X-ray-AGN to heavily obscured Compton-thick AGN with no or weak optical emission lines. On the other hand, we have unveiled a relatively large number of optical AGN without X-ray counterparts, showing the potential that surveys like ours have in detecting fainter AGN.

The optical AGN host galaxies in the OMEGA sample span a wide range in morphologies, although they are preferentially bulge-dominated disk systems. AGN also tend to prefer the more massive galaxies among the emission-line galaxy population. Interestingly, a large fraction of these objects are spiral galaxies with partially quenched star formation. We find no significant dependence of AGN activity on environment, supporting the findings of previous studies \citep{Martini_2002, Miller_2003, Martini_2006}. Note, however, that our sample spans a limited range of environments: all the OMEGA galaxies live in clusters/groups and their infall regions, and there are no true field galaxies in our sample. We therefore cannot test whether there are differences in AGN activity between field galaxies and those living in denser environments, as reported by \citet{Kauffmann_2004}.

Our analysis of the integrated star formation properties of the galaxies in the OMEGA sample shows that a large fraction of the star-forming galaxies in the A901/2 system have their star formation suppressed with respect to the field population. This quenching tends to be stronger in low-mass galaxies than in massive ones. We also confirm the existence of a significant population of intermediate- and high-mass star-forming spiral galaxies with red optical colours \citep[cf.][]{Wolf_2009}. Many of these have their star-formation suppressed while retaining their spiral morphologies.

The \Ha luminosity function of this cluster system is intermediate between that of field galaxies and the one found in more massive and denser clusters. However, within the range of environmental densities probed by our study, we do not find a strong dependence of the galaxies' star-formation activity with environment.

Because we find a relatively large population of star-forming galaxies in the A901/2 system whose star formation has been suppressed with respect to that of field galaxies, the transition from star-forming field galaxies to passive cluster galaxies has to be reasonably slow. This result agrees with the findings of \cite{Poggianti_2008}, \cite{Vulcani_2010} and \cite{Paccagnella_2016}, but contradicts the conclusions of \cite{Balogh_2004_ecology}, \cite{Verdugo_2008} and \cite{Bamford_2008}, who suggest this transition has to be relatively fast. {We attribute this discrepancy, at least partially, to differences in the definition of the star-forming galaxy samples these studies are based on and the range of environments they explore. Studies claiming that the environment affects only the fraction of star-forming galaxies but not their SSFRs tend to be based on observations that reach relatively higher SSFR limits and/or contain relatively smaller proportions of galaxies in rich clusters than studies finding also a suppression in the SSFR of cluster star-forming galaxies. The OMEGA survey is very deep, allowing the identification of galaxies with very low levels of star formation (reaching $\sim0.05\,$M$_\odot$yr$^{-1}$). Shallower surveys would consider these weakly-star-forming galaxies as passive, thus failing to identify a population of galaxies with suppressed star formation which are, nevertheless, still forming stars.

As a consequence of this slow transition, we find a relatively high number of spiral galaxies going through a phase characterized by suppressed but still significant star formation and relatively red optical colours \citep{Bamford_2009,Wolf_2009}. It is possible that these \emph{dusty red} galaxies are on their way to becoming S0s, contributing to the large population of galaxies with this morphology in clusters \citep{Dressler_1980,Calvi_2012,Fasano_2015}. However, in order to confirm this hypothesis, a detailed analysis of their structural properties, including their bulge-to-disk ratios, would be needed. It is also important to point out that because \Ha is sensitive to star formation on a very short timescale, it would be necessary to look in detail at other SFR estimators such as UV or IR light, sensitive to longer timescales, to obtain a more complete picture. This is the subject of a future paper (Wolf et al.\, in prep.)

%
The physical mechanism responsible for this transformation within the cluster environment has to be reasonably gentle, acting preferentially on the galaxies' gas but leaving their stellar disk largely intact \citep{Maltby_2010,Maltby_2012,Maltby_2015,Rodriguez_2014,Johnston_2014}. Galaxies with disturbed gas are often found in the cluster environment \citep{Yara_2011, Boesch_2013, Rawle_2014}. In fact, \citet{Boesch_2013} found that the fraction of galaxies with distorted gas kinematics is significantly higher in the A901/2 clusters than in the field, suggesting cluster-specific processes such as ram-pressure stripping \citep{Gunn_Gott_1972,Bekki_2002} as the responsible mechanism. They also found that red spirals with reduced star formation tend to have highly asymmetric rotation curves, while their stellar morphologies remain undisturbed. This implies that these galaxies are experiencing the effects of ram-pressure. 

Although ram-pressure stripping might be the main process driving the suppression of star formation in the cluster environment, other processes such as strangulation \citep{Larson_1980,Peng_2015} may affect galaxies before they reach the clusters themselves, depleting the gas reservoirs from the galaxies' halos. These processes could be responsible for the suppression of the star formation activity we have observed in many galaxies inhabiting the infall regions of the A901/2 system. Interestingly, the average SSFR of the galaxies in the cores of these clusters appears to be only slightly smaller than that of infalling galaxies, although the fraction of passive galaxies is larger in the cores. 

To disentangle all these issues, it would be useful to include velocity information in our study of this system, in the hope of constraining, at least statistically, the infall and orbital history of the galaxies. This will be the object of Weinzirl et al.\ (in prep.). Moreover, our observations allow us to produce spatially-resolved emission-line images, and thus star-formation maps, for many galaxies in our sample. The analysis of these maps will provide information on how the spatial distribution of star formation depends on intrinsic and extrinsic properties of these galaxies, and will be the focus of a future paper (Rodr\'\i guez del Pino et al., in prep.).

\section*{Acknowledgments}
Based on observations made with the Gran Telescopio Canarias (acquired through ESO Large Programme
ESO188.A-2002/GTC2002-12ESO), 
installed in the Observatorio del Roque de los Muchachos of the Instituto de Astrof\'isica de Canarias, in the island of La Palma.
We acknowledge financial support from STFC. This work has made use of The University of Nottingham HPC facility, ``Minerva". BRP acknowledges financial support from the Spanish Ministry
of Economy and Competitiveness through the Plan Nacional
de Astronom\'ia y Astrof\'isica grant AYA2012-032295.
ACS acknowledges funding from a CNPq, BJT-A fellowship (400857/2014-6). 
SPB and MEG gratefully acknowledge the receipt of an STFC Advanced Fellowship. AB is funded by the Austrian Science Foundation FWF (grant P23946-N16). This research made use of Astropy, a community-developed core Python package for Astronomy \citep{Astropy_2013} the NASA's Astrophysics Data System. Finally, we thank the anonymous referee for their constructive and useful comments and suggestions. 

\bibliographystyle{mn2e}
\bibliography{refs}
\label{lastpage}
\end{document}